\documentclass{aastex63}
\usepackage{graphicx}
\usepackage{CJK}
\acceptjournal{ApJ June 9, 2020}

\begin{document}
\begin{CJK*}{UTF8}{gbsn}

\title{On Granulation and Irregular Variation of Red Supergiants 
\footnote{Released on June, 10th, 2019}}

\correspondingauthor{Bi-Wei Jiang}
\email{bjiang@bnu.edu.cn}

\author[0000-0003-1218-8699]{Yi Ren (任逸)}
\affiliation{Department of Astronomy, Beijing Normal University, Beijing 100875, People's Republic of China}

\author[0000-0003-3168-2617]{Bi-Wei Jiang (姜碧沩)}
\affiliation{Department of Astronomy, Beijing Normal University, Beijing 100875, People's Republic of China}



\begin{abstract}
The mechanism and characteristics of the irregular variations of red supergiants (RSGs) are studied based on the RSG samples in Small Magellanic Cloud (SMC), Large Magellanic Cloud (LMC) and M31. With the time-series data from All-Sky Automated Survey for SuperNovae (ASAS-SN) and Intermediate Palomar Transient Factory survey, we use the continuous time autoregressive moving average model to estimate the variability features of the light curves and their power spectral density. The characteristic evolution timescale and amplitude of granulations are further derived from fitting the posterior power spectral density with the COR function, which is a Harvey-like granulation model. The consistency of theoretical predictions and results is checked to verify the correctness of the assumption that granulations on RSGs contribute to irregular variation. The relations between granulation and stellar parameters are obtained and compared with the results of red giant branch stars and Betelgeuse. It is found that the relations are in agreement with predictions from basic physical process of granulation and fall close to the extrapolated relations of RGB stars. The granulations in most of the RSGs evolve at a timescale of several days to a year with the characteristic amplitude of 10-1000 mmag. The results imply that the irregular variations of RSGs can be attributed to the evolution of granulations. When comparing the results from SMC, LMC and M31, the timescale and amplitude of granulation seem to increase with metallicity. The analytical relations of the granulation parameters with stellar parameters are derived for the RSG sample of each galaxy.

\end{abstract}

\keywords{stars: late-type ---
stars: oscillations (including pulsations) --- supergiants}


\section{Introduction} \label{sec:intro}
Red Supergiants (RSGs) are evolved massive stars ($M_{\rm ZAMS} \sim 9-27 M_{\odot}$), with rather low effective temperature of $\rm 3000-5000K$ \citep{2008IAUS..250...97M,2010ApJ...719.1784N,2019AandA...629A..91Y}, high luminosity of 3500-630,000$L_{\odot}$ \citep{2008IAUS..250...97M,2016ApJ...826..224M} and large radius of 100-1500$R_{\odot}$ \citep{2005ApJ...628..973L}. Their surface gravities, log $g$, range from -0.5 to 0.5.

In general, RSGs show some degree of variability in visual and infrared bands \citep{2007ApJ...667..202L,2018AandA...616A.175Y}. According to the characteristics of light curves, RSGs are divided into three types: semi-regular variables, variables with long secondary period (LSP), and irregular variables \citep{2011ApJ...727...53Y,2012ApJ...754...35Y,2019ApJS..241...35R}.
The irregular variation, present in all the three types of RSGs, becomes very popular in comparison with the bluer variables such as Cepheids or the less luminous red variables such as Miras. Meanwhile, the study of irregular variation is hindered by the difficulty to characterize the features because no simple period can be retrieved from the irregular variation. The widely accepted mechanism for irregular variation of RSGs is large convection cells firstly proposed by  \citet{1975ApJ...195..137S} since RSGs have a convective envelope due to the high opacity in the low-temperature extended atmosphere. The study suggested that the surface of  RSGs should be dominated by a few large convective cells \citep{1975ApJ...195..137S} which was later confirmed by the three dimensional simulation \citep{2002AN....323..213F}, and that the convective outer layers of RSGs cause both the spectra and the photometry to be variable \citep{2002AN....323..213F,2007A&A...469..671J}. Observationally, the $1/f$ (colored) noise component found in the power spectra means that the irregular variation may be driven by stochastic mechanisms \citep{2006MNRAS.372.1721K,2008IAUS..252..267Y} related to the large convection cells. \citet{2006MNRAS.372.1721K} find a strong 1/f noise component in almost all the stars in their sample, which suggests that convection play an important role in all the types of RSGs. With increasing power of convection, the light variation of a RSG becomes more and more irregular.

Betelgeuse ($\alpha$ Ori), as the nearest ($\sim$ 130-200 pc, see \citet{2007A&A...474..653V} and \citet{2008AJ....135.1430H}) and brightest red supergiant, has been studied in details in many aspects among which is the granulations as the observable appearance of convective cells. \citet{2009A&A...506.1351C,2010AandA...515A..12C,2011A&A...528A.120C} compared their 3D simulations of RSG convection with the interferometric observations of Betelgeuse. They concluded that the granulation pattern is detected on the surface of Betelgeuse in both the optical and the $H$ band based on the excellent fits to the observed visibility points and closure phase. They further determined the size of the convection cell to be from about 5 $\sim$ 30 mas (i.e. about 200-1200 R$_\odot$ at a distance of 200pc), which implies small number of granulations if considering the diameter of Betelgeuse to be about 1800 R$_\odot$. Other works \citep{2003RPPh...66..789M,2009A&A...508..923H,2009A&A...503..183O,2011A&A...529A.163O,2010ASPC..425..140K} also found the evidences for the existence of large convective cells in Betelgeuse.

There are only a few RSGs like Betelgeuse which can be resolved by interferometric observation thanks to very nearby location. In this sense, Betelgeuse is an exception, however the granulations of normal RSGs cannot be studied in the Betelgeuse way. Fortunately, light variation provides an alternative observable to investigate the granulations and their relation to stellar parameters. \citet{2011ApJ...741..119M} systematically analyzed the \emph{Kepler} light curves of $\sim$ 1000 red giants and successfully determined the granulation timescale  $\tau_{\rm gran}$  and power $P_{\rm gran}$. They also derived the relation of the granulation characteristics to stellar parameters. Similarly, \citet{2017MNRAS.464.1553D} analyzed the \emph{Kepler} light curve, but of a Cepheid variable V1154 Cyg, and retrieved the effective timescale of granulation to be about 3.5 days in agreement with the extrapolation of red giants. For comparison, they estimated the timescale of the granulation $\tau_{\rm gran}$ of Betelgeuse to be  $280\pm160$ days from the power density spectrum of the visual light curve assembled by the AAVSO organization, consistent with the time scale of convective motions in Betelgeuse derived from line bisector variation \citep{2008AJ....135.1450G}. These studies justified the possibility to estimate the features of RSG granulations from their light curves, and their relation to stellar parameters, which is the intention of this work.

The difference for RSGs with previous studies of red giants and Cepheid lies in the lack of the \emph{Kepler} observation, i.e. long-term continuous and accurate photometry. However, various time-domain surveys have collected large datasets, in particular, toward the nearby galaxies where RSGs are detectable for their high luminosity. In addition, extra-Galactic RSGs share the same distance which brings about the convenience to estimate stellar parameters with high relative precision. On the contrary, RSGs in our Galaxy suffer serious interstellar extinction which messes measurement of stellar distance and parameters. Following our previous study of the RSG light variation, the nearby galaxies, namely LMC, SMC, M31 and M33, are chosen as the targets. On the light variation datasets, we choose the All-Sky Automated Survey for SuperNovae (ASAS-SN, \citealt{2014ApJ...788...48S,2017PASP..129j4502K}) that covers the Small Magellanic Cloud (SMC) and the Large Magellanic Cloud (LMC), and the Intermediate Palomar Transient Factory (iPTF) survey \citep{2009PASP..121.1395L,2009PASP..121.1334R} that observed M31 and M33. We first describe selection of the RSG sample and data for light variation in Section \ref{sec:data}, then the method to derive the granulation parameters in Section \ref{sec:methods} followed by the analysis of the relation between granulation and stellar parameters in Section \ref{sec:relations}.


\section{Samples and Data} \label{sec:data}

\subsection{The preliminary sample}
The most distinguishable features of RSGs are their high luminosity and low effective temperature. Based mainly on near-infrared brightness and colors, RSGs are selected in the nearby galaxies. Historically there have been many collections of RSG samples in the MCs (e.g. \citealp{1980MNRAS.193..377F,1981MNRAS.197..385C,1983ApJ...272...99W,2000MNRAS.313..271P,2002ApJS..141...81M,2003AJ....126.2867M,2006ApJ...645.1102L,2007AJ....133.2393M,2008AJ....136.1221K,2008A&A...487.1055V,2010AJ....139.1553V}). Recently, \citet{2018AandA...616A.175Y} and \citet{2019AandA...629A..91Y,2020arXiv200510108Y} selected 773 and 1405 RSG candidates in LMC and SMC respectively by combining most of the available infrared observations and checking carefully the brightness and colors, which is the largest RSG sample in the MCs up to date. For M31 and M33, we adopted the RSG samples in our previous work  \citep{2019ApJS..241...35R}, which contain 420 and 717 RSGs in M31 and M33 respectively. \citet{2019ApJS..241...35R} find that the minimum luminosity of RSGs in M31 and M33 is about one magnitude above the theoretical limit of RSGs corresponding to $9M_\odot$. This is understandable because the observations are limited in particular to the relatively distant M31 and M33. Besides, the selection criteria are not uniform. Judging from the number of RSGs in the samples, i.e. 773, 1405, 420 and 717 RSGs in LMC, SMC, M31 and M33 respectively, we can tell that the samples are incomplete. Nevertheless, they are the largest and seemingly most reliable RSG samples in these galaxies. In addition, the samples though incomplete provide many cases to study the granulation and irregular variation in RSGs. Therefore we take them as the initial samples.

\subsection{The time-series data}

As mentioned earlier, the time-series data are taken from the ASAS-SN survey for RSGs in LMC and SMC, and from the iPTF survey for RSGs in M31 and M33. The ASAS-SN consists of 24 telescopes all around the globe, which observed variable objects in SMC and LMC at a cadence of 4-5 days for 1600 days approximately \citep{2014ApJ...788...48S,2017PASP..129j4502K}. 
ASAS-SN has magnitude limits of $g \sim 18$ and $V\sim17.3$ and photometric precision of 0.08 mag at $V \sim 16$ \citep{2018MNRAS.477.3145J}. The cadence, duration and sensitivity all fit the purpose of studying the irregular variation of RSGs in the Magellanic Clouds. The data are retrieved with a 3\arcsec-radius positional match.  For M31 and M33, the iPTF survey monitored for about 2000 days in $g$- and $R$-band, reaching 20.5 mag at 5$\sigma$ level \citep{2009PASP..121.1395L,2009PASP..121.1334R}. With a distance modulus of about 24.5 mag, this sensitivity should be compatible with the brightness of RSGs if there is not much dimming by interstellar extinction. Since \citet{2019ApJS..241...35R} already performed photometry of the RSGs in M31 and M33 from the iPTF images, their results are adopted.
However, it is noted in the later analysis that the data for RSGs in M33  leads to very uncertain determination of the granulation timescale, which may be caused by insufficient sampling since the light curves of RSGs in M33 have a similar time span as M31 but only one third of the observational points. The RSG sample of M33 is thus discarded in further study and will have to wait for better data collection. In practice, the $g$-band photometry is much less sampled and usually of lower quality in comparison with $V$-band or $R$-band due to the red color of RSG and the facility sensitivity skewed to longer wavelength, therefore the time-series data to be used are from the $V$-band observation for SMC and LMC, and the $R$-band observation for M31.

\subsection{The photometric data}

The stellar parameters, mainly the luminosity and effective temperature, are derived from multi-band photometry. We take the Two Micron All Sky Survey (2MASS) \citep{2006AJ....131.1163S} measurement in the $J$, $H$, $K_{\mathrm{S}}$ bands.
According to the study of \citet{2019ApJS..241...35R}, 18 RSGs in M31 with less than 250 photometric data points are classified as undefined type (U-type), and discarded in following analysis because the observational data are not enough for the study of variation. Finally, we collect 128, 385 and 359 light curves of RSGs in SMC, LMC and M31, respectively. Since the granulation signal could be detected in all categories of RSGs,  all the RSGs with sufficient time-series data are analyzed further, no matter whether they have been classified into semi-regular variables, irregular variables or variables with LSP \citep{2011ApJ...727...53Y,2012ApJ...754...35Y,2019ApJS..241...35R}.

\section{Determination of the granulation parameters} \label{sec:methods}


\subsection{The granulation model} \label{sec:granulation model}

The major parameters to characterize the granulations are the timescale $\tau_{\rm gran}$ and the amplitude $\sigma_{\rm gran}$. The common way to obtain the granulation parameters uses the background of power spectrum. \citet{1985ESASP.235..199H} constructed a simple but reasonable model to describe the granulations which is widely used, e.g. in red giants. The idea is that the autocovariance of granulation evolution over time can be approximated by an exponential decay function with a characteristic timescale $\tau_{\rm gran}$ and variance $\sigma^2$. This assumption leads to a Lorentz-profile power spectrum:
\begin{equation} \label{eq:Harvey}
    P(\nu)=\frac{4 \sigma^{2} \tau_{\mathrm{gran}}}{1+\left(2 \pi \nu \tau_{\mathrm{gran}}\right)^2}
\end{equation}
in which $P(\nu)$ is the total power at frequency $\nu$. It should be noted that the exponent in the denominator (`2' here) is changeable \citep{1985ESASP.235..199H}. Some modified Harvey functions are used to fit the background of the power spectral density (PSD), such as OCT \citep{2010MNRAS.402.2049H},  SYD \citep{2009CoAst.160...74H},  CAN  \citep{2010A&A...522A...1K}, and  A2Z \citep{2010A&A...511A..46M}. \citet{2011ApJ...741..119M} and \citet{2014A&A...570A..41K} used the the Harvey-like COR function \citep{2009A&A...508..877M,2011A&A...525L...9M} to fit the PSD of red giants and retrieve the granulation parameters from the \emph{Kepler} observations. Following them, we also use the COR function to fit the background of the PSD and obtain the granulation parameters in RSGs. The COR function takes the following form:
\begin{equation} \label{eq:COR}
    P(\nu)=\frac{4 \sigma^{2} \tau_{\mathrm{gran}}}{1+\left(2 \pi \nu \tau_{\mathrm{gran}}\right)^\alpha},
\end{equation}
where $\alpha$ is a positive parameter to be fitted that characterizes the slope of the decay. For various $\alpha$, a normalisation factor is required for Eq. \ref{eq:COR}, thus we change Eq. \ref{eq:COR} to
\begin{equation} \label{eq:CORc}
    P(\nu)=\frac{\xi \sigma^{2} \tau_{\mathrm{gran}}}{1+\left(2 \pi \nu \tau_{\mathrm{gran}}\right)^\alpha},
\end{equation}
where $\xi$ is a normalisation factor that depends on the value of $\alpha$. Moreover, from \citet{2013ApJ...767...34K}, $\xi$ is related to $\alpha$ as
\begin{equation} \label{eq:xi}
    \xi = 2\alpha\sin(\pi/\alpha).
\end{equation}

\subsection{Continuous Time Autoregressive Moving Average Models} \label{sec:continuous time autoregressive moving average models}

The ASAS-SN and iPTF surveys are ground-based which makes the sampling irregular. Moreover, there are usually gaps in the time-series data. Both the irregular sampling and gaps in data would deform the PSD when estimated directly from the light curves of irregular variation \citep{2009ApJ...698..895K,2014ApJ...788...33K}. Besides, the fitting would be affected by pulsation and noise of measurement when we use the COR model to fit the background of the PSD. In order to solve this problem, \citet{2014ApJ...788...33K} developed a flexible and scalable method, i.e. the Continuous-time AutoRegressive Moving Average (CARMA for short) model to estimate the variability features of the light curves and their PSD. CARMA($p,q$) process has been used to model X-ray variability of AGNs \citep{2009ApJ...698..895K,2011ApJ...730...52K} and X-ray binaries \citep{2000MNRAS.318..361N}, it is the first time to fit the granulation background using CARMA models. We make use of this model to compensate for the irregularly sampled data. A CARMA($p, q$) model can be defined as the solution to the stochastic differential equation:
\begin{equation} \label{eq:carma}
\begin{array}{c}{\frac{d^{p} y(t)}{d t^{p}}+\alpha_{p-1} \frac{d^{p-1} y(t)}{d t^{p-1}}+\cdots+\alpha_{0} y(t)} \\ {\quad=\beta_{q} \frac{d^{q} \epsilon(t)}{d t^{q}}+\beta_{q-1} \frac{d^{q-1} \epsilon(t)}{d t^{q-1}}+\cdots+\sigma\epsilon(t)}\end{array}
\end{equation}
in which $y(t)$ is the time-series data, $\alpha_{0}, \dots, \alpha_{p-1}$ are autoregressive coefficients, $\beta_{1}, \dots, \beta_{q}$ are moving average coefficients and $\epsilon(t)$ is Gaussian white noise with zero mean and variance 1. For a stationary stochastic process, $p<q$,  and it is necessary that the roots $r_{1}, ..., r_{p}$ of the autoregressive polynomial
\begin{equation} \label{eq:polynomial_carma}
	A(z)=\sum_{k=0}^{p} \alpha_{k} z^{k}
\end{equation}
have negative real parts.
The autocovariance function of CARMA(p,q) model at lag $t$ is:
\begin{equation} \label{eq:acf_carma}
	R_{p,q}(t)=\sigma^{2} \sum_{k=1}^{p} \frac{\left[\sum_{l=0}^{q} \beta_{l} r_{k}^{l}\right]\left[\sum_{l=0}^{q} \beta_{l}\left(-r_{k}\right)^{l}\right] \exp \left(r_{k} t\right)}{-2 \mathrm{Re}\left(r_{k}\right) \prod_{l=1, l \neq k}^{p}\left(r_{l}-r_{k}\right)\left(r_{l}^{*}+r_{k}\right)}.
\end{equation}
According to the Wiener-Khinchin theorem, the PSD of a wide-sense-stationary stochastic process is the Fourier transform of its autocovariance function. Then there is a PSD corresponding to a CARMA($p,q$) process:
\begin{equation} \label{eq:carma_psd}
    P(\nu)=\sigma^{2} \frac{\left|\sum_{j=0}^{q} \beta_{j}(2 \pi i \nu)^{j}\right|^{2}}{\left|\sum_{k=0}^{p} \alpha_{k}(2 \pi i \nu)^{k}\right|^{2}}.
\end{equation}
The details of mathematics and algorithm can be found in \citet{2014ApJ...788...33K}. In this algorithm,  the values of the parameters ($p,q$) must be chosen for time series analysis, which is usually done by using the Akaike information criterion (AIC; \citealp{akaike1992information}). The AIC is an estimator of the relative information lost of models for a given set of data, and AIC provides a method for model selection. \citet{hurvich1989regression} made a correction to AIC for small samples (hereafter AICc). In our work, the CARMA($p,q$) model that minimizes the AICc yields $p=3$, $q=0$.
Afterwards, the CARMA(3,0) parameters and the posterior PSD are calculated as the product of the likelihood function with a prior probability distribution using Markov Chain Monte Carlo (MCMC).

To specify the prior distribution, we need to build the relations between the coefficients of the CARMA(3,0) model and the physical parameters of convection (i.e. timescale $\tau$ and characteristic amplitude $\sigma$). This is done by considering a CARMA(1,0) model which is also known as the Ornstein-Uhlenbeck (O-U) process. The autocovariance function $R_{1,0}(t)$ and the PSD $P(\nu)$ of an O-U process are:
\begin{equation} \label{eq:acf_ou}
	R_{1,0}(t)=\frac{\sigma^{2}}{2 \alpha_{0}} e^{-\alpha_{0} t},
\end{equation}
\begin{equation} \label{eq:ou_psd}
    P(\nu)=\frac{(\sigma/\alpha_{0})^{2}}{1+(2\pi\nu/\alpha_{0})^{2}}.
\end{equation}
As can be seen, $P(\nu)$ is a Lorentzian function, and $R_{1,0}(t)$ decays with a timescale $\tau=1/\alpha_{0}$, which are related to the physical parameters $\tau$ and $\sigma$. Since the PSD of a CARMA(3, 0) process can be expressed as a sum of three Lorentzian functions, the convection characteristics can be deduced from the model. We set prior for each O-U process and assume a uniform prior on $\alpha_{0} = 1/\tau$.
The lower and upper bounds chosen for the uniform prior distribution of $1/\tau$ are set to $1/0.01$ and $1/2000$, which implies that the timescale of convection can range from 0.01-day to 2000-day. For characteristic amplitude $\sigma$, we set a uniform distribution on autocovariance at lag $t=0$ (see Eq. \ref{eq:acf_carma}), then $R_{3,0}(0)$ is a uniform distribution from 0 to 100 $\sigma_\mathrm{mag}^{2}$, where $\sigma_\mathrm{mag}$ is the standard deviation of the time series data.

Once the posterior distributions of coefficients for the CARMA(3,0) model are obtained, we can determine the autocovariance function (Eq. \ref{eq:acf_carma}) and posterior distribution of the PSD (Eq. \ref{eq:carma_psd}). After obtaining the coefficients for the CARMA(3,0) model and its posterior PSD, the residuals of the light curve need further analysis to evaluate the CARMA(3,0) model. The time-series data is fitted by using Kalman filter that estimates the mean and variance of the measured time series at time $t_{j}$ when the covariance of time-series data and measurements at time $t_{i} \ (i<j)$ are given. Then we analyze the fitting residuals, including the distribution of residuals and the autocorrelation function (ACF) of residuals. If the CARMA(3,0) model is correct, then the fitting residuals should have a normal distribution and we can perform Kolmogorov-Smirnov (KS) test on the residuals. In addition, the residuals $\chi_{i}$ and the squared residuals $(\chi_{i} - E(\chi_{i}))^{2}$ should form the white-noise sequence. If the residuals distribution deviates from the normal distribution or a large number of ACFs of residuals are outside the region assuming a white noise process, it means that the CARMA(3,0) model do not capture all correlation structures in the time series data.

\begin{figure}[ht!]
	\plotone{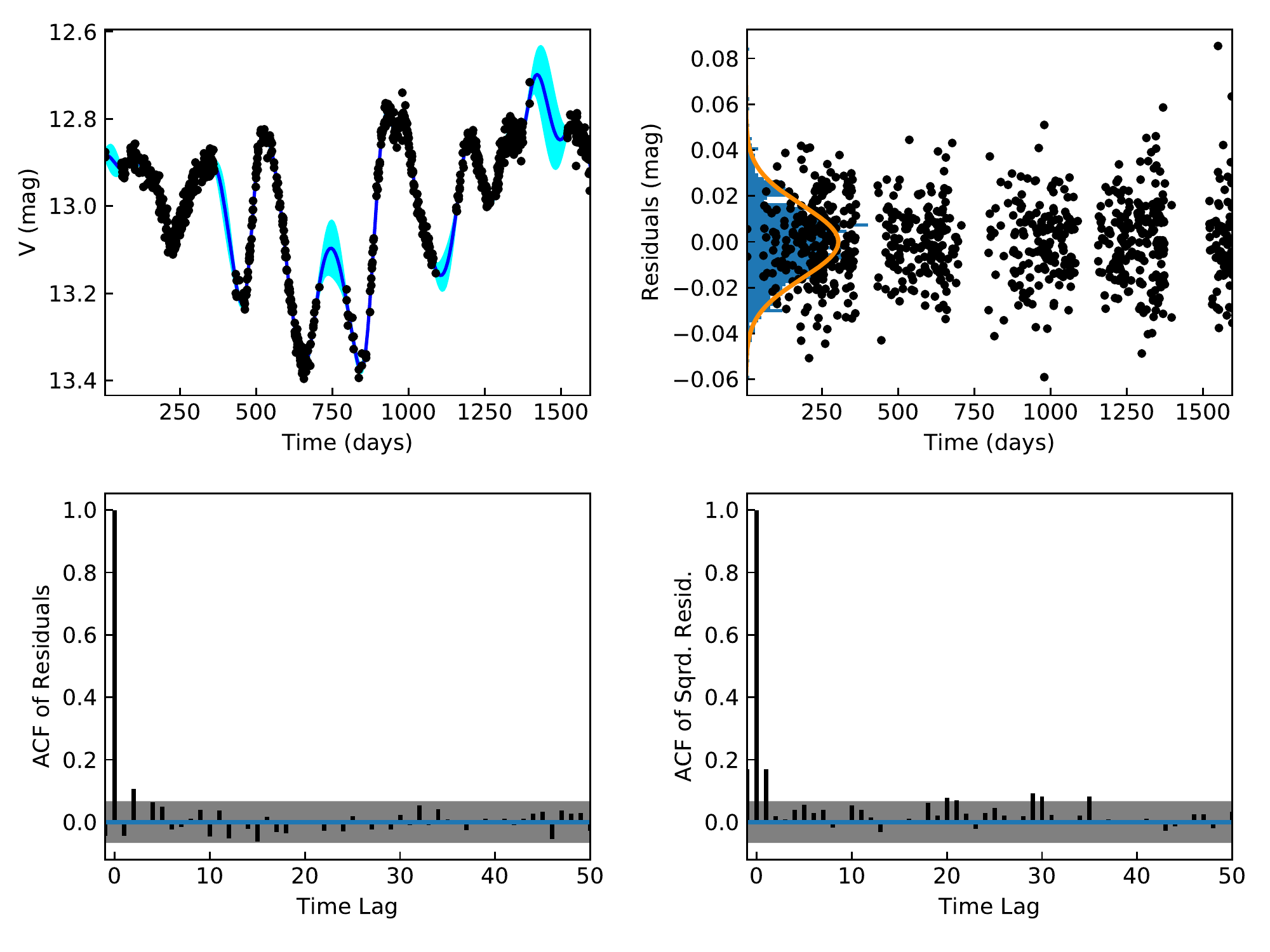}
	\caption{Example of the CARMA(3,0) model for the light curve of Source No. 180 in LMC. The left top panel is the light curve (black dots) and the fitting result (blue line) using Kalman filter with 1$\sigma$ error region of interpolated and extrapolated light curve (cyan region). The results of the residuals (black data points and blue histogram) are shown in the top right panel. The K-S test is performed on the residuals, yielding the p-value of 0.49 which is greater than the significance level (say 5\%) and cannot reject the hypothesis that the residuals come from the normal distribution (orange line). The ACFs of the residuals and their square are shown in the lower panels compared with the 95\% confidence region assuming a white noise process (gray region). \label{fig:example_lightcurve}}
\end{figure}

\begin{figure}[ht!]
	\plotone{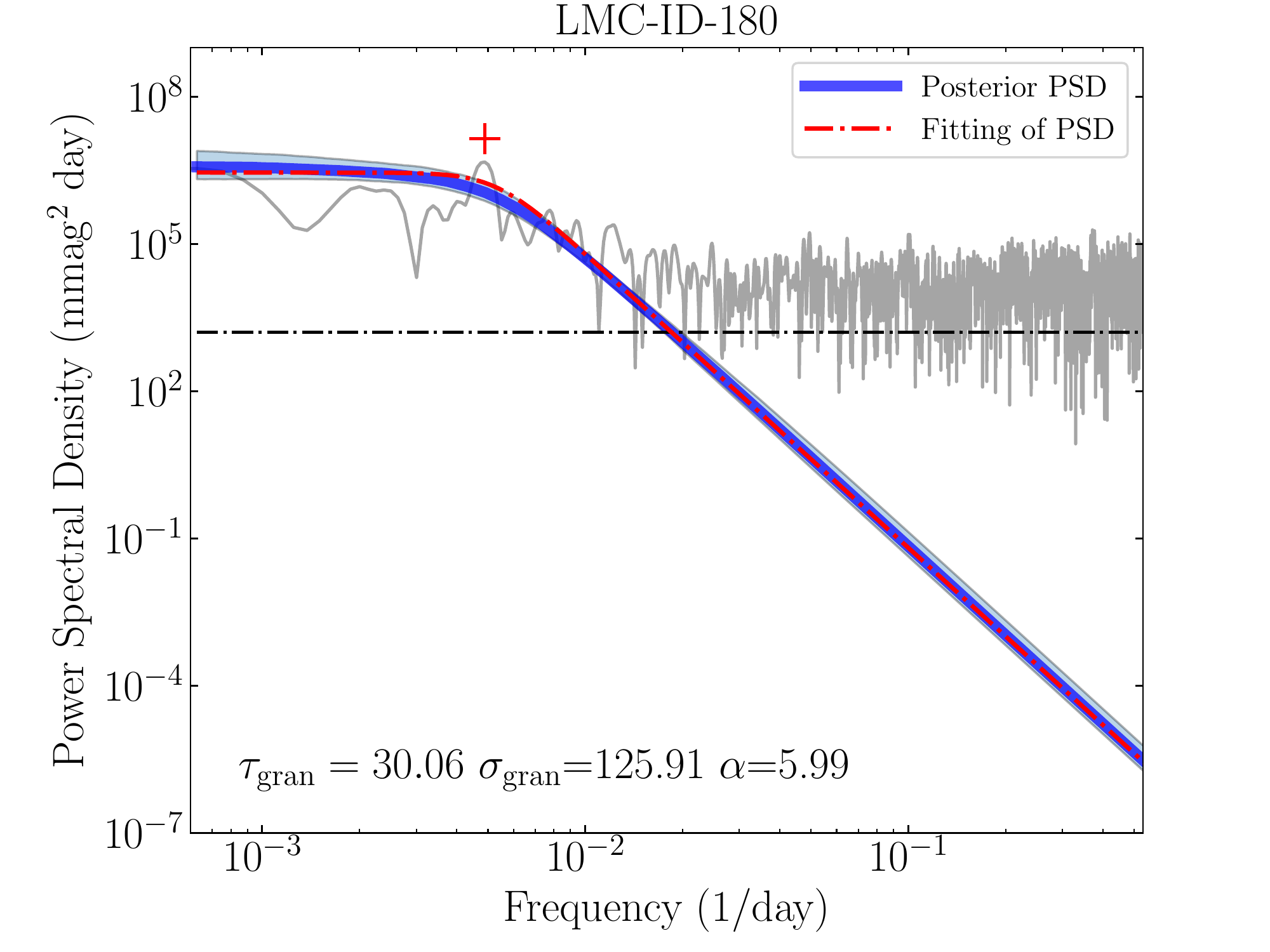}
	\caption{Example of the posterior PSD and the best fitting with blue region for the 68\% probability on PSD of Source No. 180 in LMC. Eq. \ref{eq:CORc} is used to fit the posterior PSD (blue line) from the CARMA(3,0) model. The red dashdot line shows the best fit of posterior PSD with the parameters on. The black dashdot line indicates the measurement noise level, the gray line shows the PSD based on the Lomb-Scargle (LS; \citealp{1976Ap&SS..39..447L,1982ApJ...263..835S}) method. The red cross indicates the period of pulsation.  \label{fig:example_psd}}
\end{figure}

Figure \ref{fig:example_lightcurve} and \ref{fig:example_psd} show one example (No.180 in LMC) to illustrate the method, where Figure \ref{fig:example_lightcurve} displays the result from the CARMA(3,0) model fitting to the original light curve and the residuals with its ACF expressing the goodness of fit, and Figure \ref{fig:example_psd} displays the posterior PSD of this star.


\subsection{Determination of Granulation Parameters} \label{sec:determination of granulation parameters}

The COR function is used to fit the median value of posterior PSD and obtain granulation parameters: $\tau_{\mathrm{gran}}$, $\sigma$, and the exponent $\alpha$ in Eq. \ref{eq:CORc}, which are shown in Figure \ref{fig:example_psd} for the example star.  It can be seen in Figure \ref{fig:example_psd} that the posterior PSD (blue line) traces well the background of observational PSD and is barely affected by the pulsation (red cross) and measurement noise (black dash-dot line). In order to compare characteristic timescale resulted from different $\alpha$, \citet{2011ApJ...741..119M} defined an effective timescale $\tau_{\mathrm{eff}}$ as the e-folding time of the ACF\footnote{The ACF is the inverse Fourier transform of the PSD, we calculate the ACF from PSD numerically. $\tau_{\mathrm{eff}}$ is the time required for the ACF value to reduce to $1/e$ of its initial value.}. For this purpose, the derived $\tau_{\mathrm{gran}}$ is converted to $\tau_{\mathrm{eff}}$ based on the ACF. The uncertainty of granulation parameter is calculated by the MCMC method from taking the 68\% confidence interval of the posterior PSD into account.

\begin{figure}[ht!]
	\centering
	\includegraphics[scale=0.6]{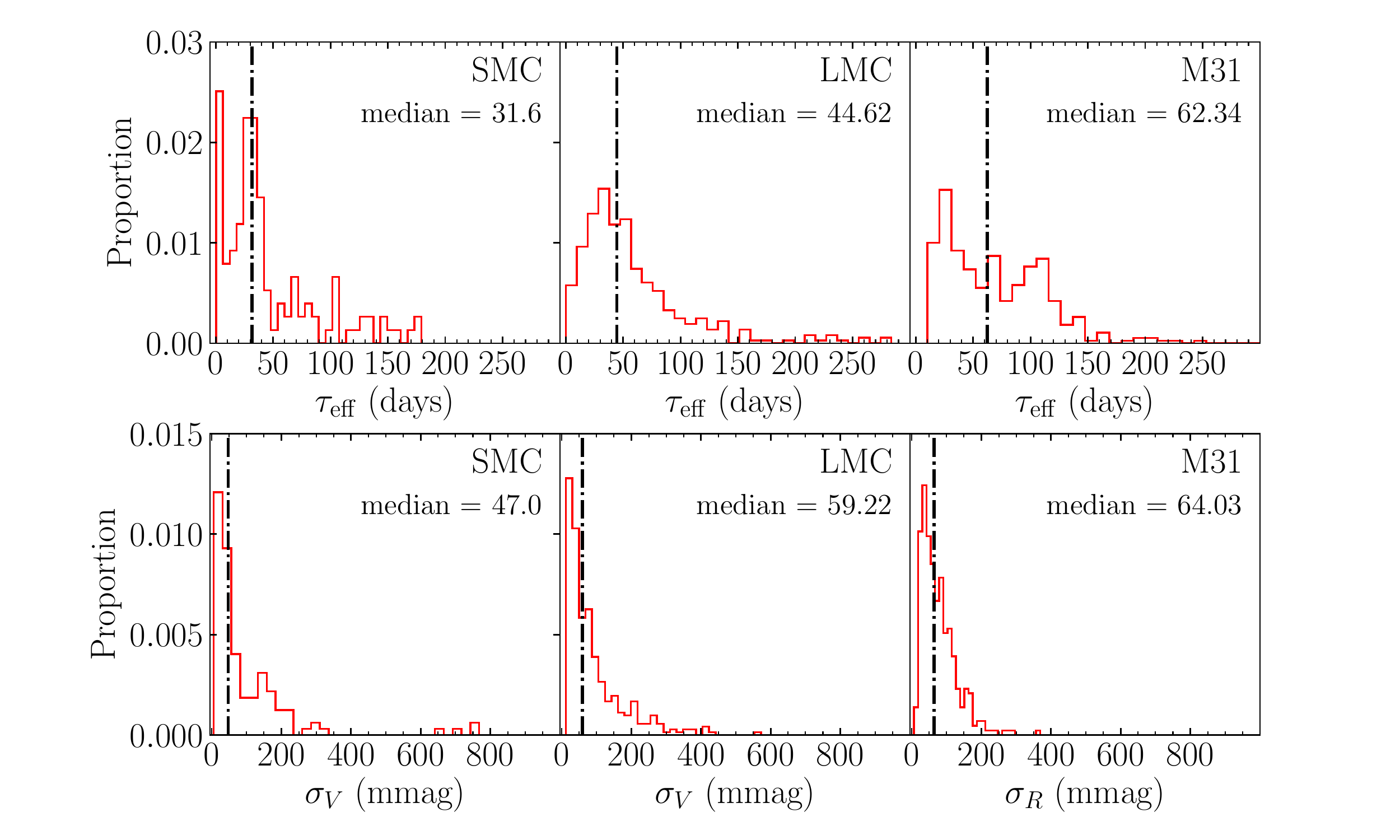}
	\caption{Distribution of the granulation parameters of RSGs in SMC (left), LMC (middle) and M31 (right), with the upper panel for granulation effective timescale $\tau_{\mathrm{eff}}$  and the lower panel for granulation amplitude $\sigma_{\mathrm{V}}$ for SMC and LMC or $\sigma_{\mathrm{R}}$ for M31.  \label{fig:distribution}}
\end{figure}

The granulation parameters of RSGs in SMC, LMC and M31 are listed in Table \ref{tab:Parameters_SMC}, \ref{tab:Parameters_LMC} and \ref{tab:Parameters_M31} and the distributions of $\tau_{\mathrm{eff}}$, $\sigma_{V}$, $\sigma_{R}$ are shown in Figure \ref{fig:distribution}. The characteristic timescale of granulations, $\tau_{\mathrm{eff}}$, ranges mainly from several days to a few hundred. This is significantly longer than in red giants where $\tau_{\mathrm{eff}}$ is generally not longer than a few days \citep{2011ApJ...741..119M}. Similarly, the characteristic amplitude of granulation, $\sigma_{\mathrm{gran}}$ ranges from several mmag to several tens mmag, also larger than in red giants \citep{2011ApJ...741..119M}. Such difference between red supergiants and red giants are understandable and will be discussed later.

The granulation parameters exhibit systematic difference between the three galaxies. The median value of $\tau_{\mathrm{eff}}$ increases from 31.60 days in SMC through 44.62 days in LMC to 62.34 days in M31. An apparent reason may be the metallicity effect since the three galaxies have very different metallicity, then $\tau_{\mathrm{eff}}$ increases with metallicity. The effect of metallicity on $\tau_{\mathrm{eff}}$ is previously observed in red giants by \citet{2017A&A...605A...3C}. They found that the time scale of granulations lengthens with metallicity for the red giants in three open clusters spanning a metallicity range from [Fe/H] $\simeq$ -0.09 to 0.32. Here the metallicity ranges from about -0.7 for SMC, through -0.46 for LMC to +0.3 for M31, $\tau_{\mathrm{eff}}$ changes by more than a factor of two. \citet{2007A&A...469..687C} found that increasing metallicity would enlarge $\sqrt{T_{\mathrm{eff}}} / g$, and $\tau_\mathrm{gran} \propto \sqrt{T_{\mathrm{eff}}} / g$ (see Section \ref{sec:basic assumptions} for details), where $T_{\mathrm{eff}}$ is effective temperature and $g$ is surface gravity, hence a long timescale. Basically, metallicity increases opacity which in turn increases the mixing length $l$, and the granulation timescale would then become longer since $\tau_{\text {gran }} \propto l / c_{s}$ as far as the sound velocity $c_{s}$ remains the same.

The median $\sigma_{\mathrm{gran}}$ changes from 47.00 mmag in SMC to 59.22 mmag in LMC both in the $V$ band, and 64.03 mmag in M31 in the $R$ band. The amplitude in the $R$ band should be converted to that in the $V$ band for comparison because the $V$ band amplitude is generally larger. There is no systematic study of the ratio of amplitude in $V$ band and $R$ band (hereafter $\mathrm{C}_{V/R}$) for RSGs. Instead, we replace with the mean ratio of red giants from a relatively large sample by \citet{2009ASPC..412..179P}, i.e. $\mathrm{C}_{V/R}=1.5$\footnote{This ratio suffers some uncertainty. For example, using the method of \citet{2019MNRAS.489.1072L}, this ratio would be around 1.2-1.3 depending on the effective temperature. The amplitude of RSGs in M31 would still be higher than the other two galaxies given this ratio, but to a smaller extent.}. Then the median $\sigma_{\mathrm{gran}}$ of RSGs in M31 should be 96.05 mmag in the $V$ band. For Betelgeuse, the $V$-band observation is available and the amplitude $\sigma_{V}$ is calculated to be 114 mmag (see Section \ref{sec:betelgeuse} for details), which is consistent with this converted value. Consequently, the characteristic amplitude of granulation increases with metallicity, which also agrees with the discovery in red giants in three open clusters with different metallicity by \citet{2017A&A...605A...3C}. Similar to the argument for the granulation timescale,  metallicity increases opacity and mixing length to affect the convective dynamics \citep{2013ApJ...767...78T}.  The increasing mixing length makes the granules occupying larger surface area \citep{2007A&A...469..687C,2013ApJ...778..117T}, and hence a higher characteristic amplitude of granulation \citep{2006A&A...445..661L,2011A&A...529L...8K}.

\begin{figure}[ht!]
	\centering
	\includegraphics[scale=0.6]{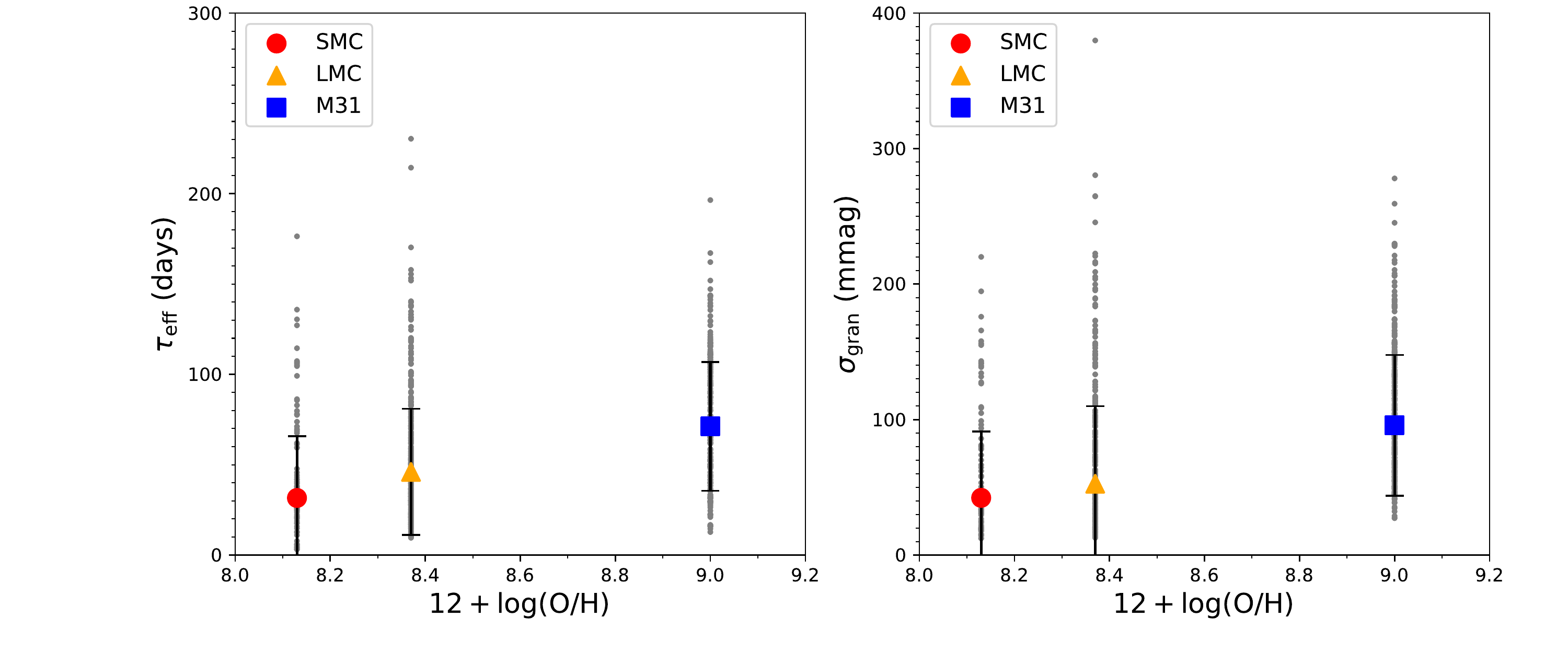}
	\caption{The granulation effective timescale (left panel) and characteristic amplitude (right panel) vs. metallicity in SMC, LMC and M31. The $12+\log(\mathrm{O/H})$ adopted for SMC, LMC and M31 is 8.13 \citep{1990ApJS...74...93R}, 8.37 \citep{1990ApJS...74...93R}, 9.00 \citep{1994ApJ...420...87Z}, respectively. The symbols indicate the median value of granulation parameters in three galaxies, and the error-bar shows $1\sigma$ range of the granulation parameters distributions. \label{fig:metallicity}}
\end{figure}

Figure \ref{fig:metallicity} shows the change of granulation effective timescale and characteristic amplitude with metallicity after excluding the outliers in the granulation parameters (c.f. Section \ref{sec:elimination of outliers} and \ref{sec:determination of the scaling relations}), where the median and 1-sigma values of $\tau$ and $\sigma_{V}$ are displayed for each galaxy at the given metallicity. It can be seen from Figure \ref{fig:metallicity} that the granulation effective timescale and characteristic amplitude increase systematically with metallicity. Meanwhile, the dispersion is significant, which could be caused by the dispersion of metallicity of individual RSGs. Moreover, the granulation parameters depends not only on metallicity, but also on stellar parameters. The influence of metallicity would be more clear after the effects of other stellar parameters such as mass and age are peeled off.

\subsection{Betelgeuse} \label{sec:betelgeuse}

\begin{figure}[ht!]
	\gridline{\fig{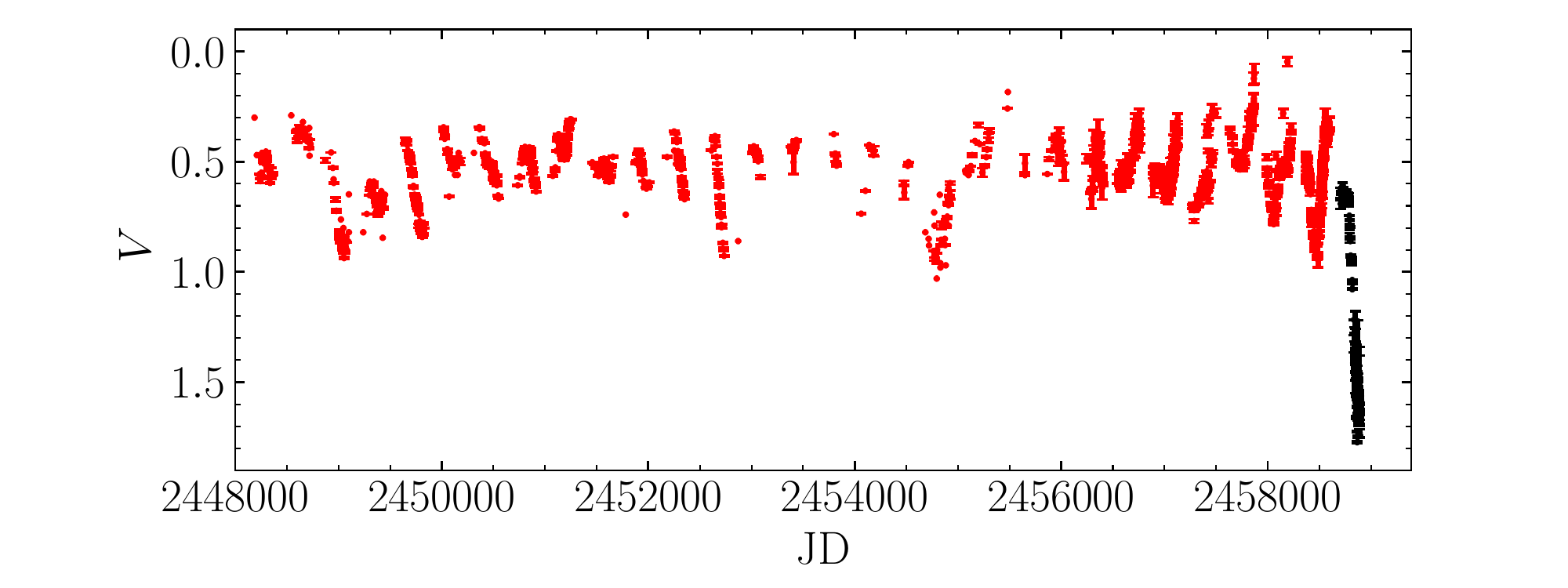}{0.9\textwidth}{(a)}
			  }
	\gridline{\fig{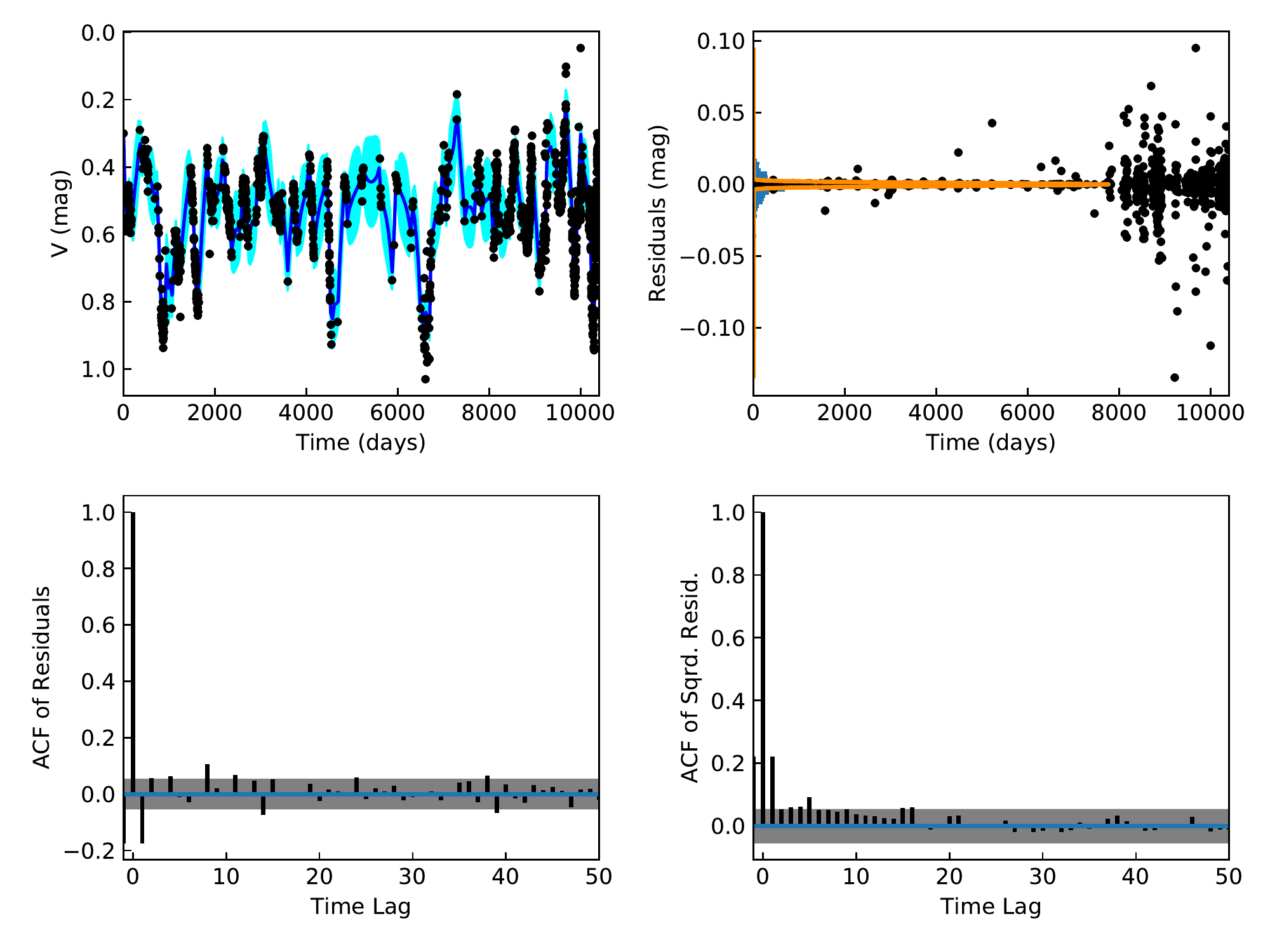}{0.9\textwidth}{(b)}
			  }
	\caption{The Betelgeuse case, (a) The visula lightcurve from AAVSO, where the data in black from the recent mysterious dimming is dropped in analysis, (b) Same as Figure \ref{fig:example_lightcurve}, but for Betelgeuse. \label{fig:betelgeuse_fit}}
\end{figure}

Betelgeuse is appended to the sample of RSGs in the three nearby galaxies as one of the nearest Galactic RSG for comparison as well as a check of the method. The timescale of granulation on Betelgeuse is derived to be several weeks to one year in optical or several years in the $H$ band by comparing the 3D radiative-hydrodynamics (RHD) simulation with interferometric observations  \citep{2010AandA...515A..12C}, and 280$\pm$160 days by fitting the ACF of the light curve \citep{2017MNRAS.464.1553D}. By analyzing the images through spectropolarimetry, \citet{2018A&A...620A.199L} find that individual convective structures can be tracked over one year but the shape and position of the convection cells change from several weeks to several months. All of these results indicate that the evolutionary time scale of the granulation of Betelgeuse ranges from several days to one year. Based on the $V$ band light curve from the AAVSO database, we calculated the timescale and the amplitude of granulation on Betelgeuse (Figure \ref{fig:betelgeuse_fit}). It should be noted that the data after JD-2458665 in Figure \ref{fig:betelgeuse_fit} is removed because the recent persistent fainting of Betelgeuse \citep{2019ATel13341....1G} is mysterious and very likely unrelated to granulation. Our method by fitting the posterior PSD of light curve yielded $\tau_{\mathrm{eff}}=138_{-3}^{+8}$ days, which is consistent with previous results in the optical bands. Besides, the amplitude in the V band $\sigma_{V} = 114_{-1}^{+19}$ mmag that agrees with expectation as discussed above.




\subsection{Gaussian Process Regression} \label{sec:gpr}
\begin{figure}[ht!]
	\plotone{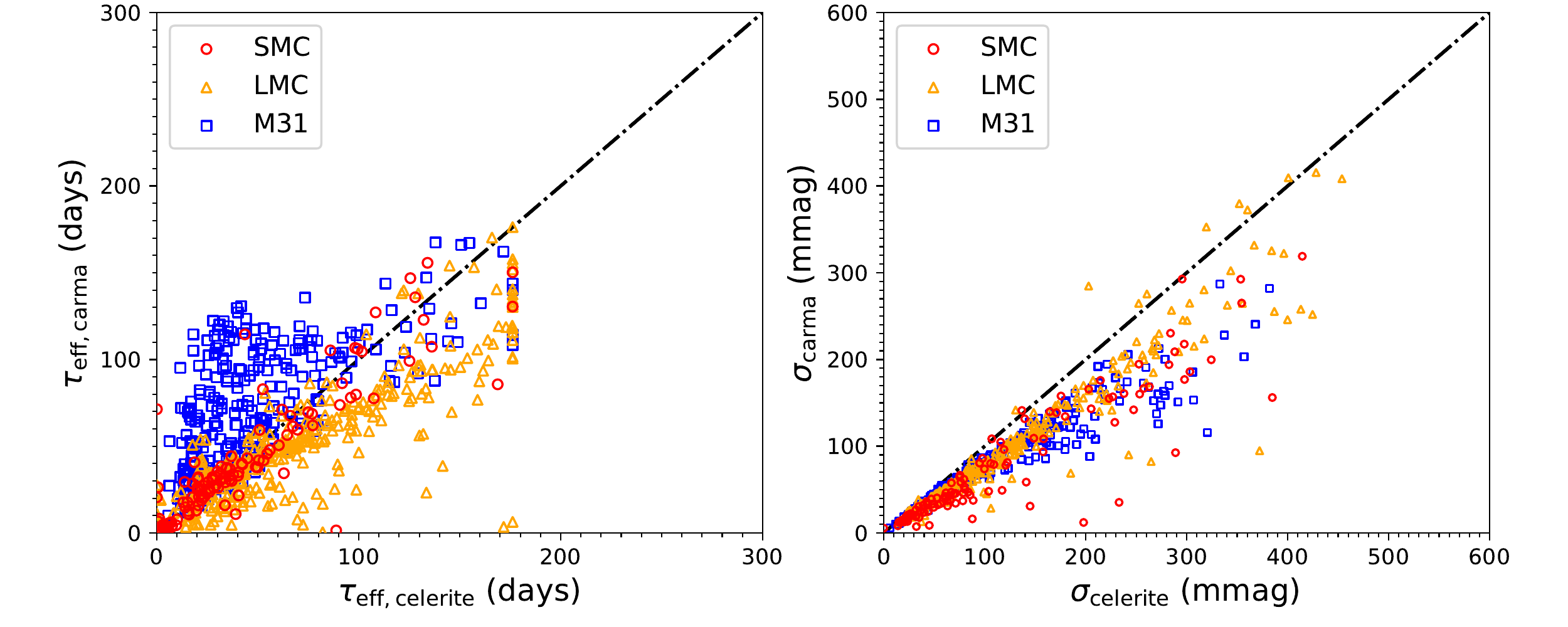}
	\caption{Comparison of the results from the CARMA and CELERITE models for  the effective timescale $\tau_\mathrm{eff}$ (left) and the characteristic amplitude $\sigma$ (right). \label{fig:carma_celerite}}
\end{figure}

Since the granulation signals can be modeled by the autocovariance and corresponding PSD (see \ref{sec:continuous time autoregressive moving average models}), the Gaussian Process (GP) can be also introduced to investigate the granulation. A key fact of GPs is that they can be completely defined by their mean and covariance functions. So the GP regression has mathematical ideas very similar to the CARMA model and has been used to model granulation and oscillations in red giant stars \citep{2019MNRAS.489.5764P}. In \citet{2019MNRAS.489.5764P}, the implementation of GP regression chosen is the Python package CELERITE \citep{2017AJ....154..220F}, a library for fast and scalable GP regression in one dimension with implementations in C++, Python, and Julia.
As indicated in \citet{2017AJ....154..220F} and \citet{2019MNRAS.489.5764P}, giving the kernel of GP as
\begin{equation} \label{eq:celerite_kernel}
	k(\tau)=S_{0} \omega_{0} \mathrm{e}^{-\frac{1}{\sqrt{2}} \omega_{0} \tau} \cos \left(\frac{\omega_{0} \tau}{\sqrt{2}}-\frac{\pi}{4}\right),
\end{equation}
the resultant PSD becomes
\begin{equation} \label{eq:celerite_psd}
	S(\omega)=\sqrt{\frac{2}{\pi}} \frac{S_{0}}{1 + \left(\omega / \omega_{0}\right)^{4}}.
\end{equation}
Since Eq. \ref{eq:celerite_psd} corresponds to the PSD of the kernel in Eq. \ref{eq:celerite_kernel} and is the same as Eq. \ref{eq:CORc} with  $\alpha=4$, this kernel can be used to capture the granulation signal in the time domain. After applying a normalization factor $K=2\sqrt{2\pi}$ \citep{2019MNRAS.489.5764P} to Eq. \ref{eq:celerite_psd} and comparing with Eq. \ref{eq:CORc}, we can derive $\tau=1/\omega_{0}$ and $\sigma=\sqrt{S_{0}\omega_{0}/\sqrt{2}}$ when $\alpha=4$. In order to compare the CARMA model with the CELERITE method, the PSD of the light curves is calculated by using CELERITE and the corresponding granulation parameters are derived. The results are also listed in Table \ref{tab:Parameters_SMC}, \ref{tab:Parameters_LMC} and \ref{tab:Parameters_M31} as a reference. Figure \ref{fig:carma_celerite} compares the results from the CARMA and CELERITE models, which shows that $\sigma_\mathrm{gran}$ agrees with each other, but $\tau_\mathrm{eff}$ presents some dispersion, especially for the RSGs in M31. This may be explained by dependence of the accuracy of $\tau_\mathrm{eff}$ on the regular sampling, while the accuracy of $\sigma_\mathrm{gran}$ does not have such dependence. In general, the two methods show consistent results since both originated from the same mathematical idea. In this work, all the further analysis are based on the results from the CARMA model.

\section{Relations Between Granulation and Stellar Parameters} \label{sec:relations}

\subsection{Basic Assumptions} \label{sec:basic assumptions}

The relations between granulation and stellar parameters are expected from both the models and observations of convection and the resultant granulation.  As verified in \citet{2009CoAst.160...74H} and \citet{2011Natur.471..608B}, the timescale of granulation is proportional to the pressure scale height and inversely proportional to the sound speed. Then the timescale of granulation can be expressed as $\tau_{\mathrm{gran}} \propto H_{p} / c_{s} \propto \sqrt{T_{\mathrm{eff}}} / g \propto \sqrt{T_{\mathrm{eff}}} R^{2} / M \propto L /M T_{\mathrm{eff}}^{3.5}$, since $H_{p} \propto T_{\mathrm{eff}} / g \text { and } c_{s} \propto \sqrt{T_{\mathrm{eff}}}$ \citep{1995A&A...293...87K}. Besides, another granulation parameter $\sigma$ can be expressed as $\sigma \propto c_{s}/\sqrt{n}$ \citep{2011A&A...529L...8K}, $\sigma$ being the rms intensity fluctuation, or granulation characteristic amplitude (hereafter $\sigma_\mathrm{gran}$) and $n$ being the number of granulations. Since $n \propto\left(R/H_{p}\right)^{2}$, $\sigma_\mathrm{gran} \propto T^{1.5}_{\mathrm{eff}} / gR \propto T^{1.5}_{\mathrm{eff}}R / M \propto L^{0.5}/M T_{\mathrm{eff}}^{0.5}$. In this work, we investigate the relations between the granulation parameters and stellar radius, effective temperature, luminosity, surface gravity and mass.

\subsection{Stellar Parameters} \label{sec:stellar parameters}
\subsubsection{Effective Temperature} \label{sec:effective temperature}

The effective temperature is calculated from the near-infrared intrinsic color index. A preparatory step is to transform the 2MASS color index $(J-K)_\mathrm{2MASS}$ into $J-K$ in the standard Johnson system \citep{1988PASP..100.1134B} following \citet{2001AJ....121.2851C}:
\begin{equation} \label{eq:11}
	J-K=\frac{\left(J-K\right)_\mathrm{2MASS}+0.011}{0.972}.
\end{equation}
The difference between the $K$ and $K_{\mathrm{2MASS}}$ bands is very small and ignored. Further, the foreground interstellar extinction is corrected to obtain the intrinsic color index $(J-K)_{0}$  by adopting $A_{K}/A_{V}=0.12$ \citep{1989ApJ...345..245C,2014ApJ...788L..12W} and $E(J-K)=A_{V}/5.79$ \citep{1998ApJ...500..525S}, in which $A_{V}=0.297$ mag for RSGs in SMC and LMC \citep{2002AJ....123..855Z,2004AJ....128.1606Z} and $A_{V}=1$ mag for RSGs in M31 \citep{2016ApJ...826..224M}.

The effective temperature $T_\mathrm{eff}$ is then calculated by its relation with $(J-K)_{0}$ \citep{2020ApJ...889...44N}  based on RSGs with known physical properties \citep{2005ApJ...628..973L,2006ApJ...645.1102L,2009ApJ...703..420M}:
\begin{equation} \label{eq:12}
	T_\mathrm{eff}=5643.5-1807.1(J-K)_{0}.
\end{equation}

\subsubsection{Luminosity and radius} \label{sec:luminosity and radius}

For RSGs in nearby galaxies, \citet{2013ApJ...767....3D} derived a function to calculate the luminosity as following:
\begin{equation} \label{eq:luminosity_mag}
\log \left(L / L_{\odot}\right)=a+b\left(m_{\lambda}-\mu\right)
\end{equation}
where $m_{\lambda}$ is the apparent magnitude in band $\lambda$ and $\mu$ is the distance modulus. We choose the $K$ band because RSGs are bright in this band and this band is little affected by interstellar extinction. The distance modulus of SMC, LMC and M31 are taken to be 18.91 \citep{2005MNRAS.357..304H}, 18.41 \citep{2006ApJ...652.1133M} and 24.40 \citep{2009A&A...507.1375P} respectively. For RSGs in SMC and LMC, \citet{2013ApJ...767....3D} obtained $a=0.90\pm0.11$ and $b=-0.40\pm0.01$ in Eq. \ref{eq:luminosity_mag}  for the extinction-corrected $K$ band magnitude. For RSGs in M31 which were not studied by \citet{2013ApJ...767....3D}, we use bolometric correction \citep{2020ApJ...889...44N} to calculate absolute magnitude and luminosity:
\begin{equation} \label{eq:14}
	\mathrm{BC}_{K}=5.567-0.757(T_{\mathrm{eff}}/1000).
\end{equation}
Thus,
\begin{equation} \label{eq:15}
	M_{\mathrm{bol}}=K_{0}+\mathrm{BC}_{K}-24.40
\end{equation}
and
\begin{equation} \label{eq:16}
	\log({L/L_{\odot}})=-0.4(M_{\mathrm{bol}}-4.75).
\end{equation}

With luminosity $L$ and $T_{\mathrm{eff}}$ calculated, the radius of RSGs can be derived from $R = (4 \pi \sigma)^{-0.5} L^{0.5} T_{\mathrm{eff}}^{-2}$. It should be noted that the errors in $L$ and $T_{\mathrm{eff}}$ are transferred into the uncertainty of $R$, consequently the error of $R$ could be much larger than $L$ and $T_{\mathrm{eff}}$ derived directly from observation.

\subsubsection{Mass and Surface Gravity} \label{sec:mass and surface gravities}
The stellar mass, $M$, is obtained from the mass-luminosity relations $L/L_{\odot}=\left(M/M_{\odot}\right)^{\gamma}$ where $\gamma \approx 4$ \citep{1971A&A....10..290S}. The surface gravity can then be calculated by $g = GM/R^2$. Again we should be careful about these secondarily derived parameters. The result depends on the accuracy of first-hand parameter $L$, the second-hand parameter $R$ as well as the mass-luminosity relation. Nevertheless, these relations have solid fundamental physics and are reasonable.

\subsection{Relation of granulation with stellar parameters}

\subsubsection{Elimination of Outliers} \label{sec:elimination of outliers}
\begin{figure}[ht!]
	\plotone{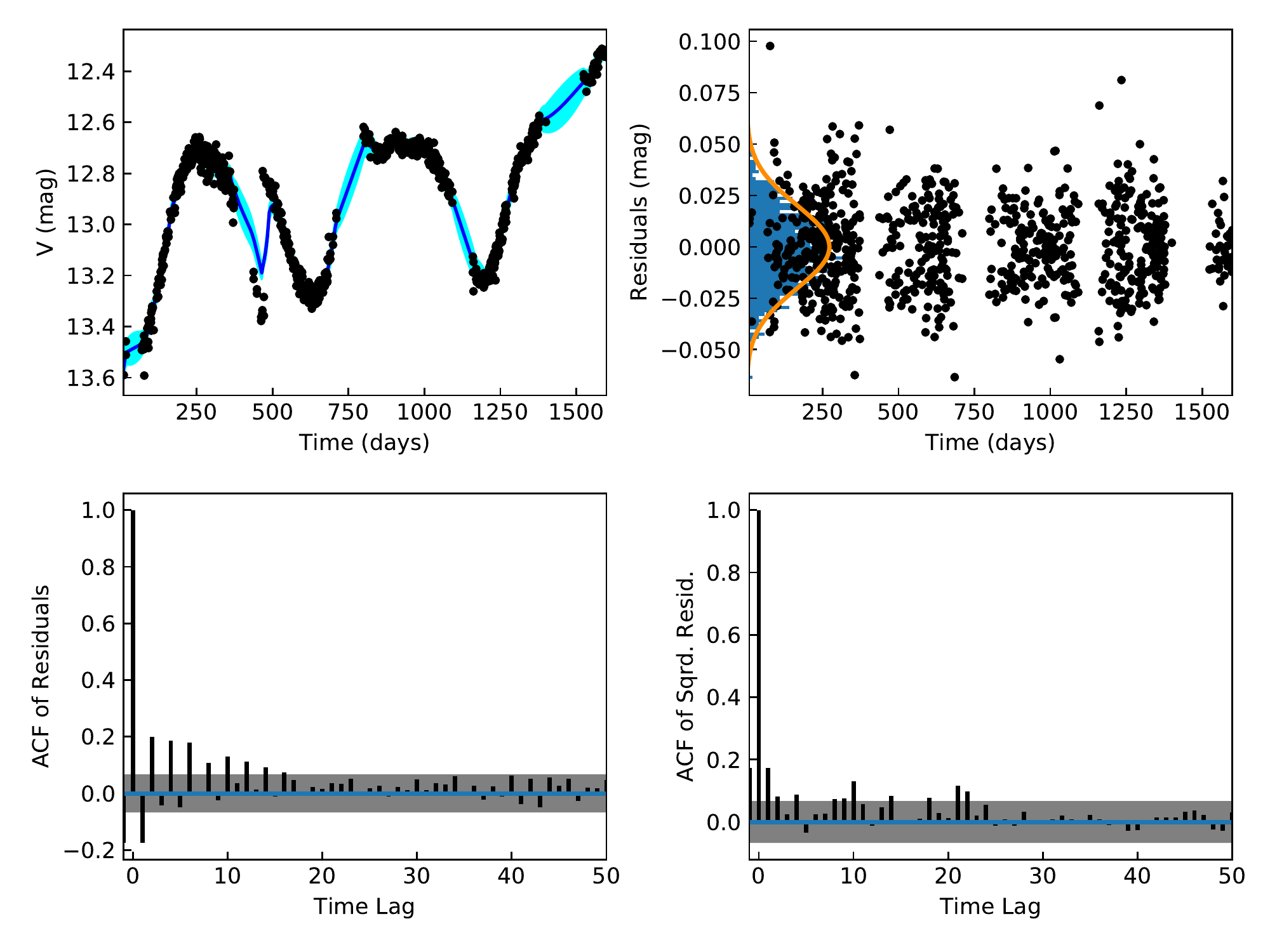}
	\caption{Same as Figure \ref{fig:example_lightcurve}, but for Source No.458 in LMC. It can be seen in the left-bottom panel that many ACFs of residuals are outside the region assuming a white noise process. It means that the CARMA(3,0) model do not capture all correlation structures in the light curve and may capture a wrong high frequency component in the light curve. \label{fig:example_outliers}}
\end{figure}

The CARMA(3,0) model may identify a faked high-frequency in light variation due to wrongly fitting the gaps in the light curve. Figure \ref{fig:example_outliers} presents an example, No.458 in LMC for which the CARMA(3.0) model captures a high-frequency variation around day 500 in the gap duration. In such case,  a very short timescale of granulation would be inferred incorrectly. The granulation parameters derived from this type of data appear as outliers to the granulation-stellar parameters relations.
The selection of more narrow boundary for uniform prior would help to avoid the occurrence of extreme values, but on the other hand, this selection of prior will force some incorrect results to fall into a reasonable range.

To reduce the effect of the outliers, the Kendall-Theil method resistant to the effects of outliers \citep{Theil1950,Sen1968} is used to calculate the coefficients for robust linear regression between granulation and stellar parameters twice. The slope of the line is the median of all possible pairwise slopes between points, meanwhile the intercept is calculated so that the line will run through the median of input data.
The points beyond the 95\% confidence region in the relations from the first fitting are regarded as outliers and dropped in the final linear fitting, which is marked as Y in the columns `Outlier$_{\rm taueff}$' and `Outlier$_{\rm sigma}$' in Table \ref{tab:Parameters_SMC}, \ref{tab:Parameters_LMC} and \ref{tab:Parameters_M31}. Actually, since the Kendall-Theil method is resistant to the effects of outliers, there is little difference between the first and second fitting parameters.

\subsubsection{Relations Between $\tau_{\mathrm{eff}}$ and Stellar Parameters} \label{sec:timescale_relations}

\begin{figure}[ht!]
	\centering
    \includegraphics[scale=0.39]{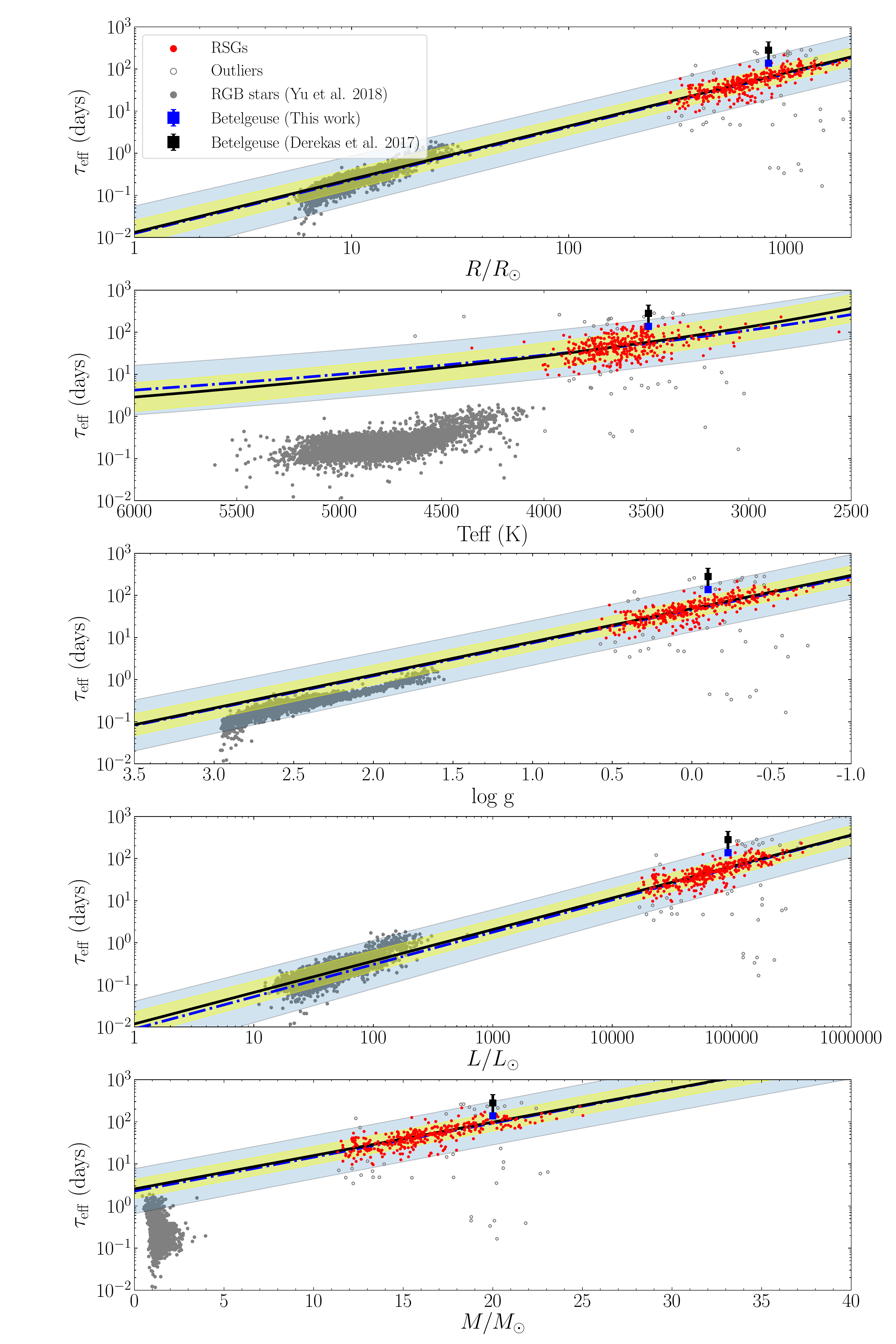}
    \caption{Relation of the granulation effective timescale $\tau_{\mathrm{eff}}$ with stellar radius, effective temperature, surface gravity, luminosity and mass (from top panel to bottom panel) for RSGs in LMC (red dots), compared with the RGB stars (gray dots) and Betelgeuse with errorbar. The blue dashdot lines are the first robust linear fit with 95\% confidence region (blue region), and the black circles are outliers beyond the 95\% confidence region from the first fitting. The black solid lines are the second robust linear fit with 95\% confidence region (yellow region). \label{fig:relations_tau_lmc}}
\end{figure}

The effective timescale $\tau_{\mathrm{eff}}$ of granulation is found to be tightly correlated with stellar parameters such as luminosity, effective temperature, radius, surface gravity and mass. The relations for RSGs in LMC is taken as an example in Figure \ref{fig:relations_tau_lmc}, where the red dots represent the RSGs to derive the linear relation and the blue shaded region marks the 95\% confidence region, while the outliers are represented by black open circles and dropped in fitting. The timescale increases with radius and luminosity, while decreases with effective temperature and surface gravity, which agree with the analysis based on the granulation assumption in Section \ref{sec:basic assumptions}. The timescale also increases with mass, which can be understood from $\tau_{\mathrm{gran}} \propto L/MT_{\mathrm{eff}}^{3.5}$ given that $L \propto M^4$. For the RSGs in SMC and M31, the relations with stellar parameters are analogous to the LMC with somewhat different coefficients, which are not present here.

Betelgeuse is compared by taking the effective timescale $\tau_{\mathrm{eff}}$ and stellar parameters from \citet{2017MNRAS.464.1553D} and \citet{2009A&A...506.1351C} as well as the timescale from this work. As can be seen from Figure \ref{fig:relations_tau_lmc}, Betelgeuse lies at the high-mass low-temperature end in comparison with the RSGs in LMC. On one hand, Betelgeuse is in agreement with the relations of $\tau_{\mathrm{eff}}$ vs. stellar parameters derived from the LMC RSGs. On the other hand, its $\tau_{\mathrm{eff}}$  seems to be larger than that at given stellar parameter in LMC. In Section \ref{sec:determination of granulation parameters}, it is found that $\tau_{\mathrm{eff}}$ increases with metallicity. Betelgeuse is a RSG in our Galaxy with [Fe/H] $\sim 0.1$ \citep{1984ApJ...284..223L} much more metal-rich than RSGs in LMC, a larger $\tau_{\mathrm{eff}}$ is expected in agreement with what is revealed here.

Since the granulation scaling relations for RSGs have never been explored before, here we only take the RGBs for comparison. Granulation is common in red giants and considered to be the mechanism for their small-amplitude light variation. The granulation effective timescale $\tau_{\mathrm{eff}}$ is calculated by \citet{2018AN....339..134D} and stellar parameters are calculated by \citet{2018ApJS..236...42Y} for a large sample of RGB stars \citep{2018ApJS..236...42Y}, which are compared with the RSGs in LMC in Figure \ref{fig:relations_tau_lmc}. It can be seen that the relations of $\tau_{\mathrm{eff}}$ with stellar radius $R$ and surface gravity log $g$ coincide with the extrapolation of the relations for RGB stars.
On luminosity, the RSGs lies slightly above but close to the extrapolation of the relation for RGBs. Meanwhile, a very good correlation is present between $\tau_{\mathrm{eff}}$ and mass for RSGs with mass range from about 10 M$_\odot$ to 20 M$_\odot$, which is not visible for RGBs within a relatively narrow range of mass.
A remarkable difference occurs in the relation with effective temperature in that the relation of $\tau_{\mathrm{eff}}$ with $T_{\mathrm{eff}}$ of RSG stars lies significantly above that of RGBs.  As mentioned above, the relation between $\tau_{\mathrm{eff}}$, surface gravity and effective temperature can be expressed as $\tau_{\text {eff}} \propto \sqrt{T_{\mathrm{eff}}} / g$, thus $\mathrm{log} \ \tau_{\text {eff}} \propto 0.5 \times \mathrm{log} \ T_{\mathrm{eff}} - \mathrm{log} \ g$. The difference of the surface gravity would lead to a vertical shift -$\mathrm{log} \ g$ in the $\tau_{\mathrm{eff}}$ vs. $\mathrm{log} \ T_{\mathrm{eff}}$ diagram. Because the $\mathrm{log} \ g$ value is $\sim$0.0 for RSGs while $\sim 2.0$ for RGBs, $\tau_{\mathrm{eff}}$ of a RSG would be  $\sim$2.0 dex larger than of a RGB star at the same  $T_{\mathrm{eff}}$, which is consistent with the difference observed in Figure \ref{fig:relations_tau_lmc}.
In a word, the relations of the granulation characteristic timescale $\tau_{\mathrm{eff}}$  are consistent with the red supergiant Betelgeuse and RGB stars, and in agreement with the granulation theory.

\subsubsection{Relations Between $\sigma_{\mathrm{gran}}$ and Stellar Parameters} \label{sec:amplitude_relations}

\begin{figure}[ht!]
	\centering
    \includegraphics[scale=0.39]{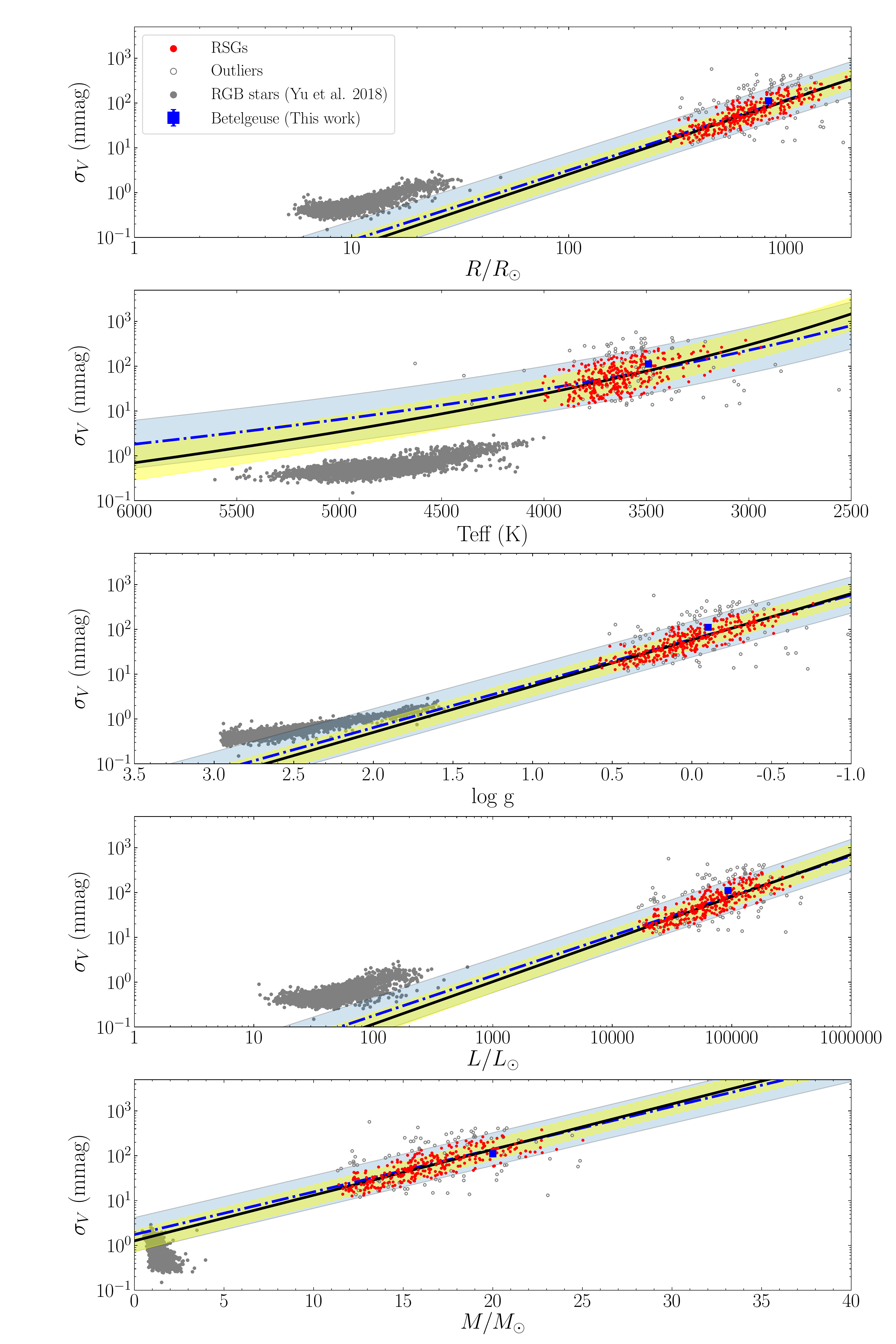}
    \caption{Same as Figure \ref{fig:relations_tau_lmc}, but for the granulation amplitude $\sigma_{\mathrm{gran}}$. \label{fig:relations_sigma_lmc}}
\end{figure}

The characteristic amplitude of granulation is expected to relate to the stellar parameters as analyzed in Section \ref{sec:basic assumptions}. In the same way as  dealing with the parameter $\tau_{\mathrm{eff}}$, we use the RSGs in LMC as an example. The results are displayed in Figure \ref{fig:relations_sigma_lmc}. The amplitude increases with stellar radius, luminosity and mass, but decreases with effective temperature and surface gravity, which agree with the analysis based on the granulation assumption in Section \ref{sec:basic assumptions}. Indeed, the relations with individual stellar parameter are not very straight. For example, the relation with effective temperature can be derived from $\sigma_\mathrm{gran} \propto  L^{0.5}/M T_{\mathrm{eff}}^{0.5}$. With $L \propto R^2 T_{\mathrm{eff}}^{4.0}$, this expression can be written as $\sigma_\mathrm{gran} \propto R T_{\mathrm{eff}}^{1.5}/M$. Whether $\sigma_{\mathrm{gran}}$ increases or decreases with $T_{\mathrm{eff}}$ would depend on the relation of $M/R$ with $T_{\mathrm{eff}}$. Assuming that $T_{\mathrm{eff}} \propto (M/R)^\beta$, it can be inferred from the inverse correlation between $\sigma_{\mathrm{gran}}$ and $T_{\mathrm{eff}}$ that the exponent $\beta < 0.67$. 

The relations of $\sigma_{\mathrm{gran}}$ with stellar parameters are compared with RGB stars in Figure \ref{fig:relations_sigma_lmc}. The deviation of RGBs from the derived relation for RSGs looks much more apparent than in the case of $\tau_{\mathrm{eff}}$ present in Figure \ref{fig:relations_tau_lmc}. That RGB stars lie below the $\sigma_{\mathrm{V}}$ vs. $T_{\mathrm{eff}}$ relation can be easily understood from the smaller surface gravity of RSGs since $\sigma_\mathrm{gran}$ inversely proportion to $g$ as $\propto T^{1.5}_{\mathrm{eff}} / gR$. The other deviations cannot be directly explained in this way, depending on the mutual relations between stellar parameters, which is beyond the scope of this paper, but we present more discussion in Section \ref{sec:offsets between RSGs and RGB stars}.



\subsection{Determination of the Scaling Relations} \label{sec:determination of the scaling relations}

\subsubsection{Scaling Relations between granulation parameters, $L$ and $T_\mathrm{eff}$} \label{sec:scaling relations L R}

The brief analysis in Section \ref{sec:basic assumptions} reveals that the granulation parameters, $\tau_{\mathrm{eff}}$ and $\sigma_{\mathrm{gran}}$ are related to multiple stellar parameters such as $L$, $T_{\mathrm{eff}}$, $R$, $\log g$, $M$ etc. Because these stellar parameters are not independent, the relations with all these parameters would be unnecessary. Although there could be various options to combine some parameters, we decide to choose $L$ and $T_{\mathrm{eff}}$ as the principle ones because only $L$ and $T_{\mathrm{eff}}$ are derived directly from the observations while the other parameters depend on $L$ and $T_{\mathrm{eff}}$ and would have more uncertainties (see Section \ref{sec:stellar parameters}). Without knowing the analytical form of the relations, a polytropic equation is adopted, i.e.  $\tau_{\mathrm{eff}} = 10^a L^{\beta} T_{\mathrm{eff}}^{\gamma} $, and $\sigma_{\mathrm{gran}}$ takes the same form with different power law indexes.

The indexes $a$,  $\beta$, and $\gamma$ of the scaling relations are determined by the MCMC method using PyMC3 \citep{salvatier2016probabilistic}. After excluding the outliers in granulation parameters, 94, 342 and 226 RSGs are contained to determine these unknowns for $\tau_{\mathrm{eff}}$ in SMC, LMC and M31 respectively, and 73, 271 and 188 RSGs for $\sigma_{\mathrm{gran}}$. For RSGs in SMC, we notice that 7 stars (No. 3839, 6999, 11765, 19164, 20019, 38653, 43337) deviate from the scaling relation significantly. We re-checked their light curves which were found to be inconsistent with RSGs. A further examination found that the 7 objects are relatively blue and faint, with four of them being classified as Cepheid variables \citep{2018A&A...616A...1G,2018A&A...618A..30H}. Thus they are discarded in final fitting. This case analysis implies that there may be a potential to identify a RSG from the consistency with the scaling relation or not.

\begin{figure}[ht!]
	\gridline{\figtight{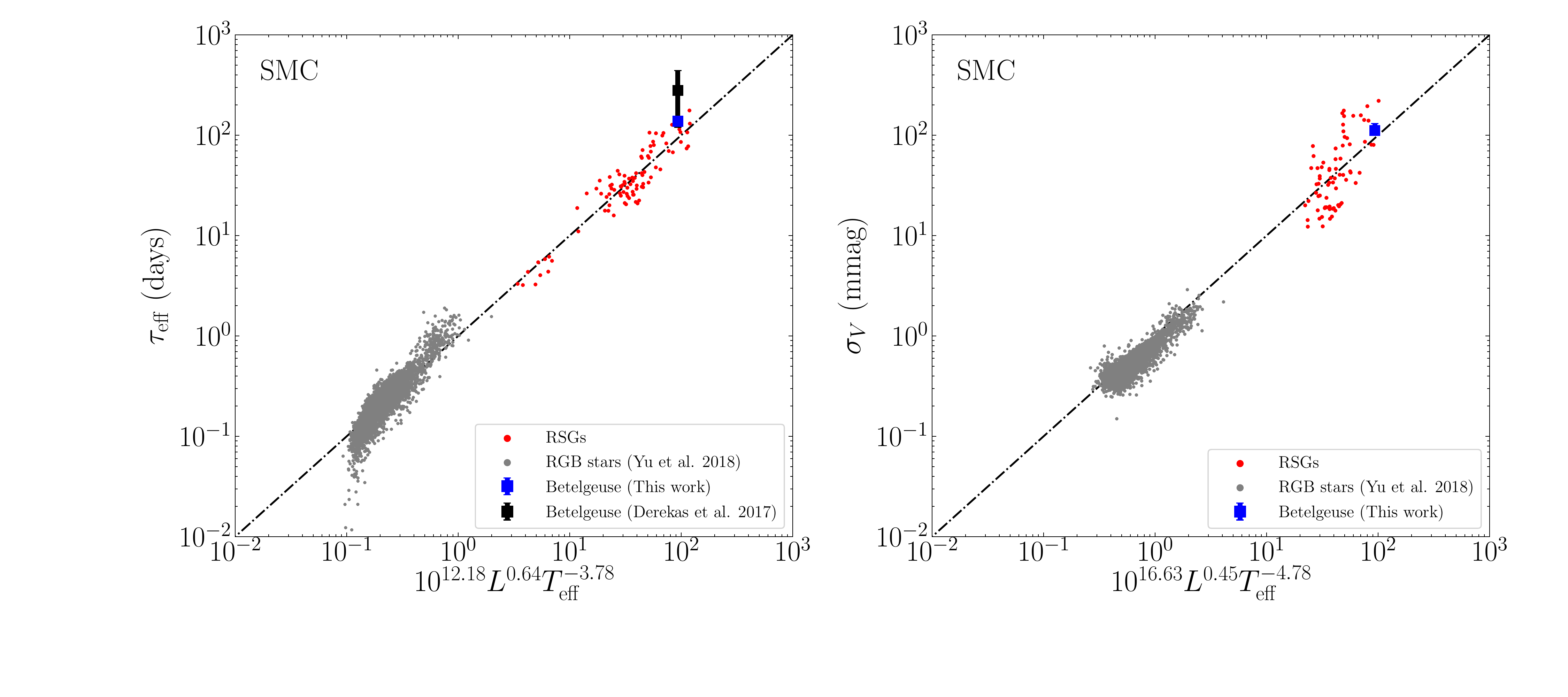}{0.95\textwidth}{(a)}}
	\gridline{\figtight{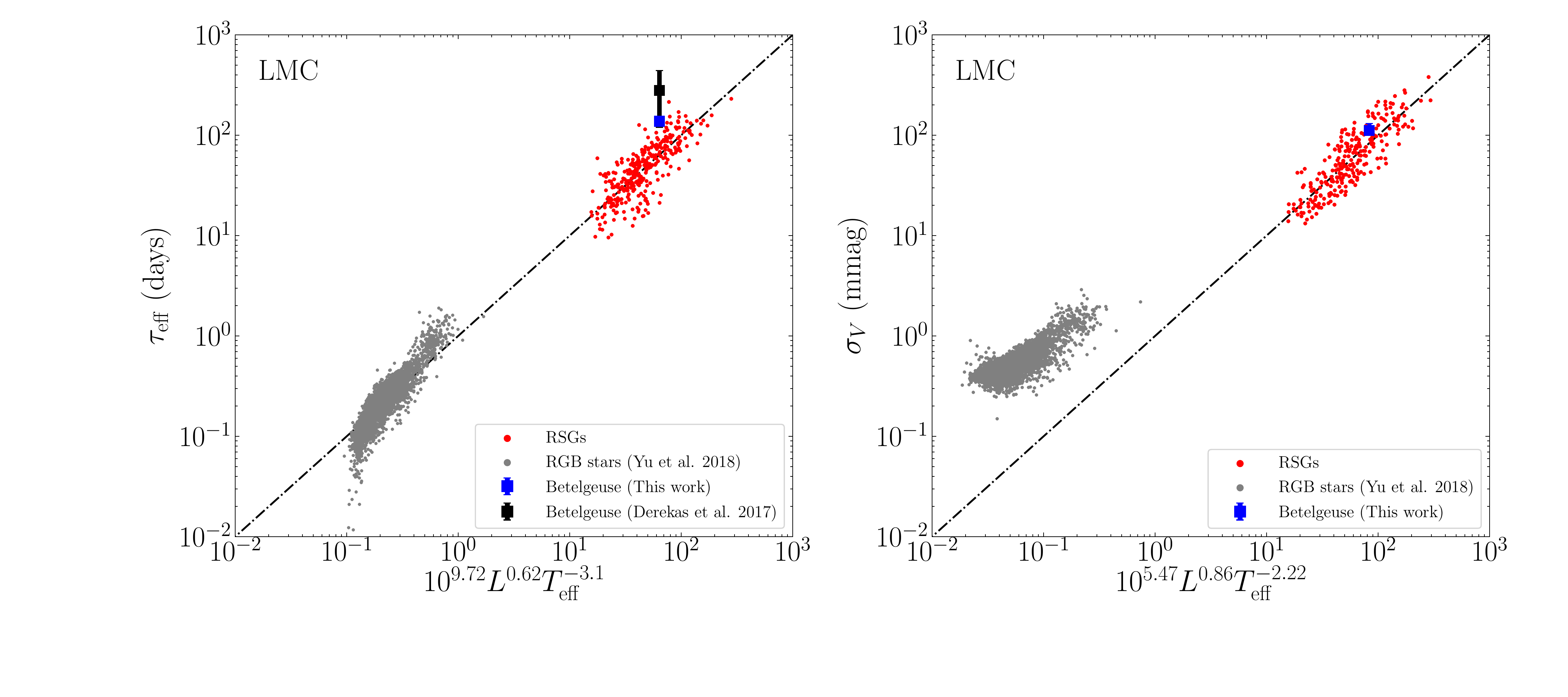}{0.95\textwidth}{(b)}}	
	\gridline{\figtight{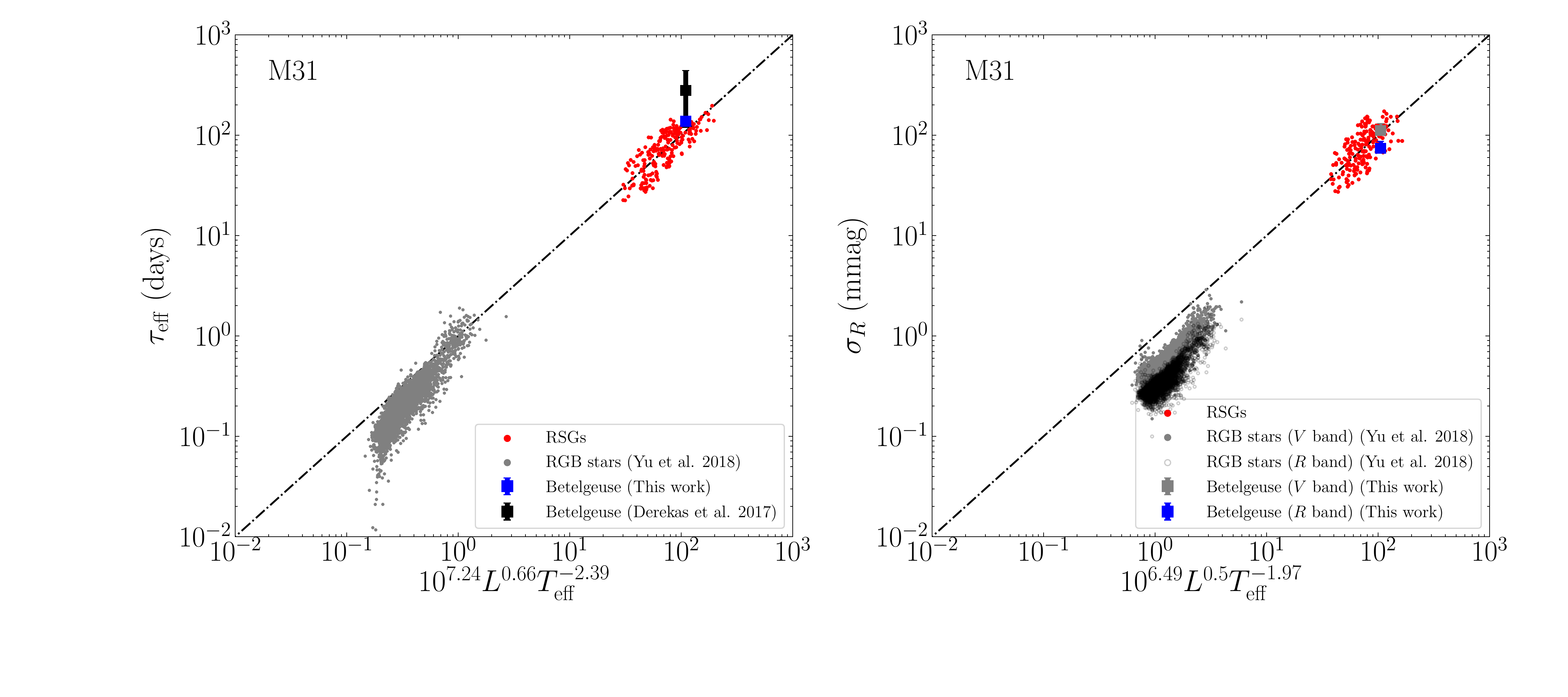}{0.95\textwidth}{(c)}}
	\caption{Scaling relations of the granulation effective timescale $\tau_{\mathrm{eff}}$ (left) and amplitude $\sigma_{\mathrm{gran}}$ (right) with stellar luminosity and effective temperature for RSGs in SMC (top), LMC (middle) and M31 (bottom). For RSGs in M31, $\sigma_{V}$ of RGB stars (gray dots) and Betelgeuse (gray square) is scaled down to $\sigma_{R}$ by dividing by 1.5. \label{fig:relations}}
\end{figure}

The fitted indexes are listed in Table \ref{tab:relations_tau} for $\tau_{\mathrm{eff}}$ and Table \ref{tab:relations_sigma} for $\sigma_{\mathrm{gran}}$ together with the correlation coefficient. Figure \ref{fig:relations} illustrates the scaling relations of the granulation parameters with stellar luminosity and effective temperature for RSGs in SMC, LMC and M31 respectively. The granulation amplitudes of RGBs were derived from the $Kepler$ or $CoRoT$ observation whose effective wavelength of filters is analogous to the $V$ band, thus their $\sigma_{\mathrm{gran}}$ is divided by a factor of 1.5 when comparing with $\sigma_{\mathrm{R}}$ of RSGs in M31. It can be seen that the correlations are apparent in all the cases, for both $\tau_{\mathrm{eff}}$ and $\sigma_{\mathrm{gran}}$, and for all the three galaxies. Among them, the relation for $\tau_{\mathrm{eff}}$  in M31 is relatively loose, which can be understood by presence of bigger gaps and less regular sampling in the time-series data from iPTF which leads to the more uncertain determination of $\tau_{\mathrm{eff}}$. Besides, the coefficients $\beta_{\sigma},\gamma_{\sigma}$ in the scaling relations for SMC and LMC ($\sigma_{V}$) are larger than for M31 ($\sigma_{R}$). In other words, $\sigma_{V}$ is more sensitive to stellar parameters than $\sigma_{R}$, because the intensity depends on the opacity at given wavelength. The metallicity effect is visible in comparison with the same RGB stars sample. The relative position of RGBs moves from slightly above the fitting line to beneath, i.e. $\tau_{\mathrm{eff}}$ and $\sigma_{\mathrm{gran}}$ systematically rise from SMC through LMC to M31. This effect is explained in Section \ref{sec:determination of granulation parameters}.

From the fundamental analysis in Section \ref{sec:basic assumptions}, $\tau_{\mathrm{eff}} \propto L/(MT_\mathrm{eff}^{3.5})$ and $\sigma \propto L/(MT_\mathrm{eff}^{0.5})$, then the relations can be expressed with only $L$ and $T_\mathrm{eff}$ if $M$ is approximated by $L^{1/4}$, i.e. $\tau_{\mathrm{eff}} \propto L^{0.75}T_\mathrm{eff}^{-3.5}$ and $\sigma \propto L^{0.75}T_\mathrm{eff}^{-0.5}$. In comparison with this fundamental analysis, the fitted relations with $L$ coincide with the expectation in that $\beta$ ranges from about 0.4 to 0.9 for both $\tau_{\mathrm{eff}}$ and $\sigma_\mathrm{gran}$. On the other hand, the fitted relations with $T_\mathrm{eff}$ show some discrepancy. The fitted value of $\gamma$ for $\tau_{\mathrm{eff}}$ ranges from about -2 to -4, more or less in agreement with the expected -3.5 though slightly different. Meanwhile, the value of $\gamma$ for $\sigma_\mathrm{gran}$ ranges from about -2 to -5 significantly different from -0.5, the expected value, which means $\sigma_\mathrm{gran}$ is much more sensitive to $T_\mathrm{eff}$ than expected. In Section \ref{sec:offsets between RSGs and RGB stars}, we give a possible explanation for this large discrepancy.

\subsubsection{Offsets between RSGs and RGB stars} \label{sec:offsets between RSGs and RGB stars}

\begin{figure}[ht!]
	\centering
	\includegraphics[scale=0.5]{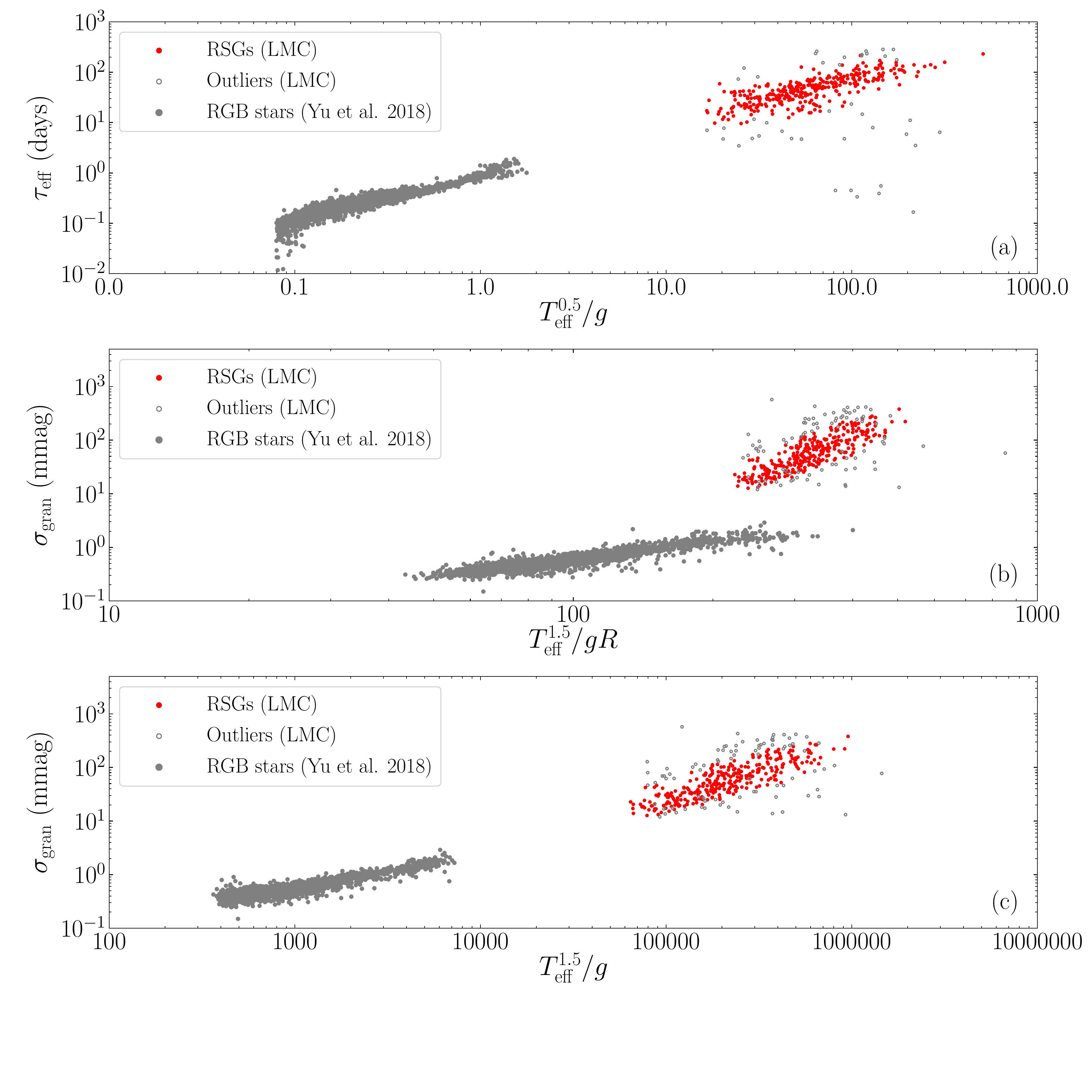}
    \caption{Granulation parameters vs. multiple stellar parameters: (a) $\tau_{\mathrm{eff}}$ vs. $T_{\mathrm{eff}}^{0.5}/g$, (b) $\sigma_{R}$ vs. $T_{\mathrm{eff}}^{1.5}/gR$, and (c) $\sigma_{R}$ vs. $T_{\mathrm{eff}}^{1.5}/g$.  \label{fig:new_scaling}}
\end{figure}

The relations of granulation parameters with single stellar parameter exist offsets between RSGs and RGB stars, which can be seen in Figure \ref{fig:relations_tau_lmc} and \ref{fig:relations_sigma_lmc}. In particular, the relation with  effective temperature exhibits the most significant offset. According to the analysis of Section \ref{sec:timescale_relations}, the offset of $\tau_\mathrm{eff}$ with $T_\mathrm{eff}$ could be attributed to the difference in surface gravity. In order to clarify the effect of $g$, we plot $\tau_\mathrm{eff}$ versus $\sqrt{T_\mathrm{eff}}/g$ in Figure \ref{fig:new_scaling} (a) for the RSGs in LMC and the RGB stars by following the fundamental relation of granulation  $\tau_\mathrm{eff} \propto \sqrt{T_\mathrm{eff}}/g$ . The offsets between RSGs and RGBs disappear in the $\tau_\mathrm{eff}$ versus $\sqrt{T_\mathrm{eff}}/g$ diagram, which confirms previous judgement.

The case of the characteristic amplitude  $\sigma_\mathrm{gran}$ is more complicated as discussed in Section \ref{sec:amplitude_relations}. Following the basic assumption of \citet{2011A&A...529L...8K}, $\sigma \propto c_{s}/\sqrt{n}$, then $\sigma_\mathrm{gran} \propto T_\mathrm{eff}^{1.5}/gR$. Figure \ref{fig:new_scaling} (b) presents the relation of $\sigma_\mathrm{gran}$ with  $\propto T_\mathrm{eff}^{1.5}/gR$ for the RSGs in LMC and the RGB stars, but the offset still exists between RSGs and RGBs, which may indicate the $\sigma_\mathrm{gran} \propto T_\mathrm{eff}^{1.5}/gR$ relation is incorrect. From the very fundamental concept of granulation,  the characteristic amplitude is an indicator of the granulation cell energy, which should be proportional to the mixing length of convection and the sound speed. Taking that mixing length is on the order of pressure scale height, then $\sigma_\mathrm{gran} \propto c_{s}H_{p} \propto T_\mathrm{eff}^{1.5}/g \propto T_\mathrm{eff}^{1.5}R^{2}/M \propto L/T_\mathrm{eff}^{2.5}M$. Figure \ref{fig:new_scaling} (c) plot the relation between $\sigma_\mathrm{gran}$ and $T_\mathrm{eff}^{1.5}/g$, which does bring about the agreement between RSGs and RGBs. Moreover, substituting the relation between $L$ and $M$ yields $\sigma_\mathrm{gran} \propto L^{0.75}/T_\mathrm{eff}^{2.5}$, a very different relation with $T_\mathrm{eff}$ in comparison with previous works. This roughly coincides with the scaling relation found in previous section \ref{sec:scaling relations L R}, where $\sigma_\mathrm{gran}$ is found to be proportional to $T_\mathrm{eff}^{(-2) - (-5)}$.


\section{Summary and Conclusion} \label{sec:summary and conclusion}

The irregular variation and its relation with granulation of RSGs are analyzed for the RSG samples in SMC, LMC and M31 with the time series data from ASAS-SN and iPTF.  The CARMA(3,0) model is used to capture the correlation structures of the 128, 385, 359 light curves of RSGs in SMC, LMC and M31 respectively, which forms the basis to use the COR method to fit the posterior PSD and obtain the characteristic evolution timescale $\tau_{\mathrm{eff}}$ and amplitude $\sigma_{\mathrm{gran}}$ of granulations. It is found that the granulations in most of the RSGs evolve at a timescale of several days to a year with the characteristic amplitude of 10-1000 mmag. Both $\tau_{\mathrm{eff}}$ and $\sigma_{\mathrm{gran}}$ increases from SMC through LMC to M31, which may indicate the influence of metallicity on the granulation characteristics.

The relations between granulation and stellar parameters are analyzed. Both the effective timescale $\tau_{\mathrm{eff}}$ and amplitude $\sigma_{\mathrm{gran}}$ of granulations are tightly correlated with stellar effective temperature, luminosity, radius, surface gravity and mass. These relations agree with the expectations from the granulation model. In addition, they accord with the extrapolation of the relations derived from the \emph{Kepler} light curves of red giant branch stars. The numerals of the  Galactic red supergiant Betelgeuse whose granules were detected by interferometric observations support these relations as well. A scaling relation is derived for the granulation parameters with stellar luminosity and effective temperature for the RSG sample in SMC, LMC and M31 respectively.

The results of the ranges of granulation parameters and their relation with stellar parameters agree with the assumption that irregular variation of RSGs can be attributed to the evolution of granulations on their surface. The large granulation characteristic amplitude implies that the surface of RSGs are dominated by a few but huge granules.

\acknowledgments This paper is published to commemorate the 60th anniversary of the Department of Astronomy, Beijing Normal University. We are grateful to Drs. Ming Yang and Yong Shi, and Ms. Yuxi Wang for very helpful discussions, and the anonymous referee for very good suggestions. This work is supported by the National Natural Science Foundation of China through the project NSFC 11533002.
This work has made use of data from the surveys by ASAS-SN and iPTF as well as the AAVSO database.

%

\vspace{5mm}
\facilities{}


\software{Astropy \citep{2013A&A...558A..33A},
		  SExtractor \citep{1996A&AS..117..393B},
		  PyMC3 \citep{salvatier2016probabilistic},
		  TOPCAT \citep{2005ASPC..347...29T},
		  LMfit \citep{2018zndo...1699739N}
          }




\bibliography{paper}{}
\bibliographystyle{aasjournal}



\begin{figure}[ht!]
	\plotone{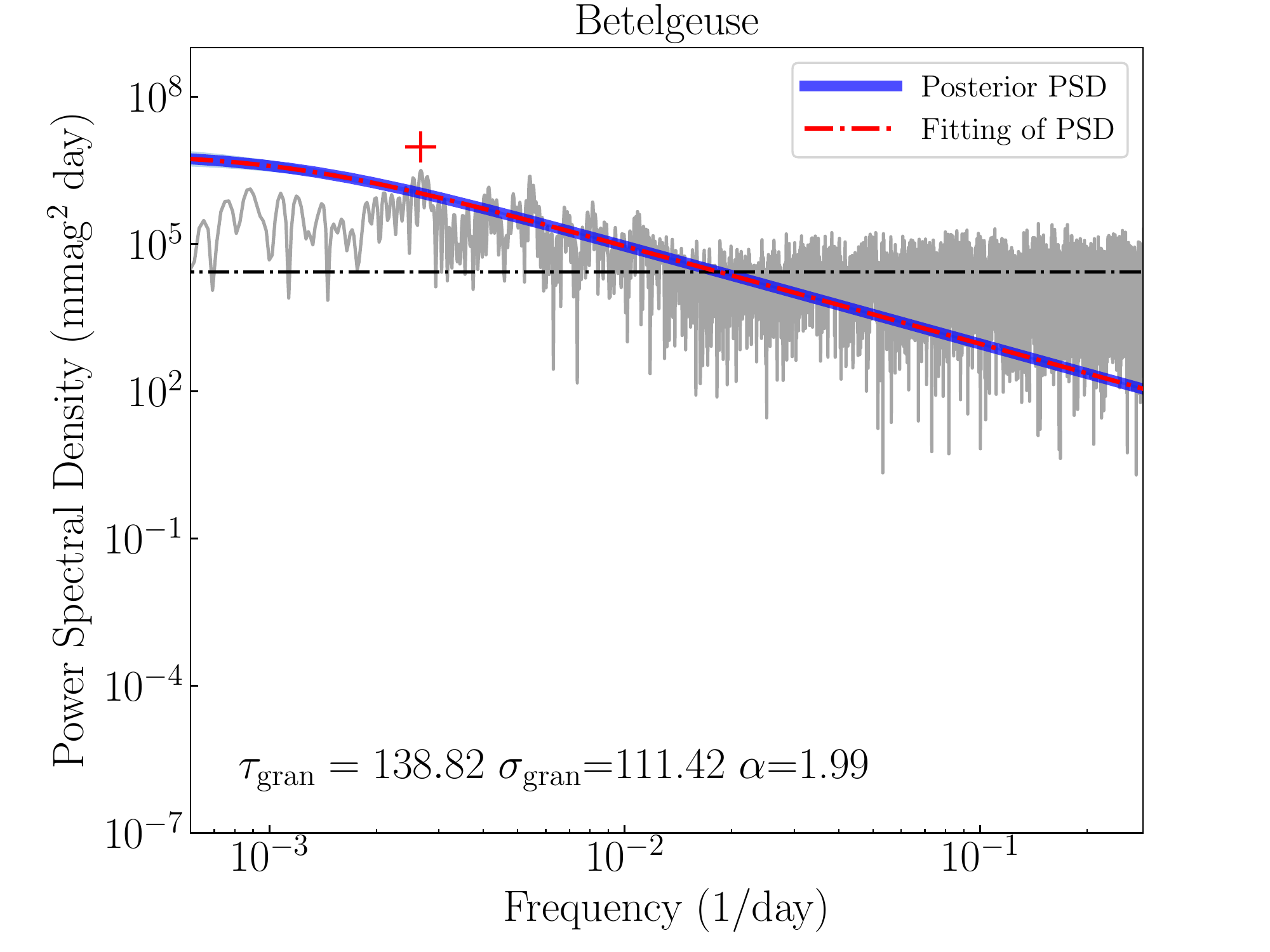}
	\caption{Same as Figure \ref{fig:example_psd}, but for Betelgeuse. \label{fig:betelgeuse_pdf}}
\end{figure}

\begin{longrotatetable}
	\begin{deluxetable*}{ccccccccccccccc} \label{tab:Parameters_SMC}
	\tablecaption{Granulation and Stellar Parameters of RSGs in SMC}
	\tablewidth{0pt}
	\tablehead{
	\colhead{Galaxy\_ID$^\mathrm{a}$} & \colhead{$L$} & \colhead{$M$} & \colhead{$T_\mathrm{eff}$} & \colhead{$R$} & \colhead{$\mathrm{log}\ g$} & \colhead{$\tau$$^\mathrm{b}$} & \colhead{$\alpha$$^\mathrm{c}$} & \colhead{$\sigma$$^\mathrm{d}$} & \colhead{${\tau_\mathrm{eff}}$$^\mathrm{e}$} & \colhead{$\tau_\mathrm{celerite}$$^\mathrm{f}$} & \colhead{$\sigma_\mathrm{celerite}$$^\mathrm{g}$} & \colhead{$\tau_\mathrm{eff,celerite}$$^\mathrm{h}$} & \colhead{Outlier\_taueff} & \colhead{Outlier\_sigma} \\
	\colhead{} & \colhead{($L_{\odot}$)} & \colhead{($M_{\odot}$)} & \colhead{(K)} & \colhead{($R_{\odot}$)} & \colhead{} & \colhead{(day)} & \colhead{} & \colhead{(mmag)} & \colhead{(day)} & \colhead{(day)} & \colhead{(mmag)} & \colhead{(day)} & \colhead{} & \colhead{}
	}
	\startdata
	SMC\_ID\_768 & 9826 & 10.0 & 3693 & 243 & 0.7 & $7.0_{-1.95}^{+1.14}$ & $3.0_{-0.03}^{+0.01}$ & $66.54_{-0.64}^{+0.63}$ & $10.63_{-2.96}^{+1.73}$ & 9.14 & 75.51 & 15.98 & Y & N \\
	SMC\_ID\_1349 & 16291 & 11.3 & 3739 & 305 & 0.5 & $15.93_{-0.1}^{+0.11}$ & $6.12_{-0.0}^{+0.01}$ & $24.72_{-0.27}^{+0.27}$ & $31.39_{-0.2}^{+0.22}$ & 11.57 & 43.93 & 20.27 & N & N \\
	SMC\_ID\_3182 & 30334 & 13.2 & 3504 & 473 & 0.2 & $34.21_{-0.29}^{+0.37}$ & $4.26_{-0.03}^{+0.01}$ & $93.56_{-1.08}^{+1.41}$ & $61.35_{-0.52}^{+0.66}$ & 38.43 & 157.78 & 67.41 & N & N \\
	SMC\_ID\_3544 & 24051 & 12.5 & 3493 & 424 & 0.3 & $2.63_{-0.05}^{+0.02}$ & $3.71_{-0.0}^{+0.01}$ & $108.58_{-0.34}^{+0.43}$ & $4.45_{-0.08}^{+0.03}$ & 5.46 & 106.94 & 9.59 & Y & N \\
	SMC\_ID\_3839 & 1884 & 6.6 & 5074 & 56 & 1.8 & $2.22_{-0.01}^{+0.03}$ & $7.49_{-0.02}^{+0.02}$ & $35.25_{-0.31}^{+0.37}$ & $4.55_{-0.02}^{+0.06}$ & 1.66 & 233.03 & 2.9 & N & N
	\enddata
	\tablecomments{$^\mathrm{a}$The Galaxy\_ID are the same as \citet{2019AandA...629A..91Y}. \\
	$^\mathrm{b}$Granulation timescale estimated by CARMA(3,0) model. \\
	$^\mathrm{c}$Exponent in Eq. \ref{eq:CORc}. \\
	$^\mathrm{d}$Characteristic amplitude of granulation estimated by CARMA(3,0) model. \\
	$^\mathrm{e}$Granulation effective timescale estimated by CARMA(3,0) model. \\
	$^\mathrm{f}$Granulation timescale estimated by CELERITE model. \\
	$^\mathrm{g}$Characteristic amplitude of granulation estimated by CELERITE model. \\
	$^\mathrm{h}$Granulation effective timescale estimated by CELERITE model. \\
	(This table is available in its entirety in machine-readable form.)}
	\end{deluxetable*}
\end{longrotatetable}

\begin{longrotatetable}
	\begin{deluxetable*}{ccccccccccccccc} \label{tab:Parameters_LMC}
	\tablecaption{Granulation and Stellar Parameters of RSGs in LMC}
	\tablewidth{0pt}
	\tablehead{
	\colhead{Galaxy\_ID$^\mathrm{a}$} & \colhead{$L$} & \colhead{$M$} & \colhead{$T_\mathrm{eff}$} & \colhead{$R$} & \colhead{$\mathrm{log}\ g$} & \colhead{$\tau$$^\mathrm{b}$} & \colhead{$\alpha$$^\mathrm{c}$} & \colhead{$\sigma$$^\mathrm{d}$} & \colhead{${\tau_\mathrm{eff}}$$^\mathrm{e}$} & \colhead{$\tau_\mathrm{celerite}$$^\mathrm{f}$} & \colhead{$\sigma_\mathrm{celerite}$$^\mathrm{g}$} & \colhead{$\tau_\mathrm{eff,celerite}$$^\mathrm{h}$} & \colhead{Outlier\_taueff} & \colhead{Outlier\_sigma} \\
	\colhead{} & \colhead{($L_{\odot}$)} & \colhead{($M_{\odot}$)} & \colhead{(K)} & \colhead{($R_{\odot}$)} & \colhead{} & \colhead{(day)} & \colhead{} & \colhead{(mmag)} & \colhead{(day)} & \colhead{(day)} & \colhead{(mmag)} & \colhead{(day)} & \colhead{} & \colhead{}
	}
	\startdata
	LMC\_ID\_3 & 28651 & 13.0 & 3674 & 419 & 0.3 & $6.26_{-0.33}^{+0.2}$ & $4.73_{-0.06}^{+0.03}$ & $26.5_{-0.26}^{+0.27}$ & $11.64_{-0.61}^{+0.37}$ & 10.92 & 29.48 & 19.19 & Y & N \\
	LMC\_ID\_4 & 50626 & 15.0 & 3690 & 551 & 0.1 & $18.19_{-0.16}^{+0.21}$ & $4.8_{-0.05}^{+0.02}$ & $37.53_{-0.43}^{+0.59}$ & $33.85_{-0.3}^{+0.39}$ & 24.85 & 62.86 & 43.58 & N & N \\
	LMC\_ID\_5 & 23353 & 12.4 & 3597 & 394 & 0.3 & $20.66_{-1.07}^{+0.98}$ & $2.83_{-0.04}^{+0.03}$ & $13.6_{-0.14}^{+0.12}$ & $30.14_{-1.56}^{+1.43}$ & 18.65 & 16.52 & 32.71 & N & Y \\
	LMC\_ID\_8 & 23439 & 12.4 & 3854 & 344 & 0.5 & $5.69_{-0.05}^{+0.05}$ & $6.77_{-0.03}^{+0.03}$ & $19.16_{-0.14}^{+0.13}$ & $11.44_{-0.1}^{+0.1}$ & 4.65 & 30.02 & 8.17 & N & N \\
	LMC\_ID\_10 & 30164 & 13.2 & 3752 & 412 & 0.3 & $7.29_{-0.07}^{+0.08}$ & $6.13_{-0.02}^{+0.02}$ & $28.72_{-0.13}^{+0.15}$ & $14.33_{-0.14}^{+0.16}$ & 18.56 & 33.53 & 32.53 & N & N
	\enddata
	\tablecomments{$^\mathrm{a}$The Galaxy\_ID are the same as \citet{2018AandA...616A.175Y}. \\
	$^\mathrm{b}$Granulation timescale estimated by CARMA(3,0) model. \\
	$^\mathrm{c}$Exponent in Eq. \ref{eq:CORc}. \\
	$^\mathrm{d}$Characteristic amplitude of granulation estimated by CARMA(3,0) model. \\
	$^\mathrm{e}$Granulation effective timescale estimated by CARMA(3,0) model. \\
	$^\mathrm{f}$Granulation timescale estimated by CELERITE model. \\
	$^\mathrm{g}$Characteristic amplitude of granulation estimated by CELERITE model. \\
	$^\mathrm{h}$Granulation effective timescale estimated by CELERITE model. \\
	(This table is available in its entirety in machine-readable form.)}
	\end{deluxetable*}
\end{longrotatetable}

\begin{longrotatetable}
	\begin{deluxetable*}{ccccccccccccccc} \label{tab:Parameters_M31}
	\tablecaption{Granulation and Stellar Parameters of RSGs in M31}
	\tablewidth{0pt}
	\tablehead{
	\colhead{Galaxy\_ID$^\mathrm{a}$} & \colhead{$L$} & \colhead{$M$} & \colhead{$T_\mathrm{eff}$} & \colhead{$R$} & \colhead{$\mathrm{log}\ g$} & \colhead{$\tau$$^\mathrm{b}$} & \colhead{$\alpha$$^\mathrm{c}$} & \colhead{$\sigma$$^\mathrm{d}$} & \colhead{${\tau_\mathrm{eff}}$$^\mathrm{e}$} & \colhead{$\tau_\mathrm{celerite}$$^\mathrm{f}$} & \colhead{$\sigma_\mathrm{celerite}$$^\mathrm{g}$} & \colhead{$\tau_\mathrm{eff,celerite}$$^\mathrm{h}$} & \colhead{Outlier\_taueff} & \colhead{Outlier\_sigma} \\
	\colhead{} & \colhead{($L_{\odot}$)} & \colhead{($M_{\odot}$)} & \colhead{(K)} & \colhead{($R_{\odot}$)} & \colhead{} & \colhead{(day)} & \colhead{} & \colhead{(mmag)} & \colhead{(day)} & \colhead{(day)} & \colhead{(mmag)} & \colhead{(day)} & \colhead{} & \colhead{}
	}
	\startdata
	M31\_ID\_15 & 26843 & 12.8 & 3563 & 431 & 0.3 & $32.81_{-1.04}^{+0.63}$ & $3.8_{-0.03}^{+0.02}$ & $84.42_{-0.78}^{+0.8}$ & $56.34_{-1.79}^{+1.08}$ & 29.95 & 107.08 & 52.52 & N & N \\
	M31\_ID\_16 & 69501 & 16.2 & 4087 & 527 & 0.2 & $44.88_{-0.7}^{+0.58}$ & $3.39_{-0.02}^{+0.01}$ & $63.61_{-0.5}^{+0.53}$ & $73.09_{-1.14}^{+0.94}$ & 25.68 & 74.99 & 45.04 & N & N \\
	M31\_ID\_17 & 129351 & 19.0 & 3948 & 770 & -0.1 & $44.76_{-0.53}^{+1.0}$ & $5.88_{-0.02}^{+0.07}$ & $125.04_{-2.05}^{+2.53}$ & $87.57_{-1.04}^{+1.96}$ & 65.79 & 159.7 & 115.32 & N & N \\
	M31\_ID\_18 & 37507 & 13.9 & 3459 & 540 & 0.1 & $18.73_{-0.23}^{+0.21}$ & $3.42_{-0.02}^{+0.02}$ & $25.99_{-0.18}^{+0.19}$ & $30.69_{-0.38}^{+0.34}$ & 13.98 & 32.68 & 24.5 & Y & Y \\
	M31\_ID\_19 & 35527 & 13.7 & 3537 & 503 & 0.2 & $35.09_{-0.36}^{+0.48}$ & $2.93_{-0.01}^{+0.01}$ & $48.02_{-0.29}^{+0.37}$ & $52.56_{-0.54}^{+0.72}$ & 12.8 & 55.6 & 22.42 & N & N
	\enddata
	\tablecomments{$^\mathrm{a}$The Galaxy\_ID are the same as \citet{2019ApJS..241...35R}. \\
	$^\mathrm{b}$Granulation timescale estimated by CARMA(3,0) model. \\
	$^\mathrm{c}$Exponent in Eq. \ref{eq:CORc}. \\
	$^\mathrm{d}$Characteristic amplitude of granulation estimated by CARMA(3,0) model. \\
	$^\mathrm{e}$Granulation effective timescale estimated by CARMA(3,0) model. \\
	$^\mathrm{f}$Granulation timescale estimated by CELERITE model. \\
	$^\mathrm{g}$Characteristic amplitude of granulation estimated by CELERITE model. \\
	$^\mathrm{h}$Granulation effective timescale estimated by CELERITE model. \\
	(This table is available in its entirety in machine-readable form.)}
	\end{deluxetable*}
\end{longrotatetable}

\begin{deluxetable*}{cccccccc}[ht] \label{tab:relations_tau}
	\tablecaption{Fitted model parameters of $\tau_{\mathrm{eff}}$ with stellar luminosity and effective temperature in $\tau_{\mathrm{eff}} \propto 10^{a_{\tau}} L^{\beta_{\tau}} T^{\gamma_{\tau}}_\mathrm{eff}$.}

	\tablewidth{0pt}
	\tablehead{
		\colhead{Galaxy} & \colhead{$a_{\tau}$} & \colhead{$\beta_{\tau}$} & \colhead{$\gamma_{\tau}$} & \colhead{Correlation} \\
	}
	\startdata
	SMC & $12.18\pm1.55$ & $0.64\pm0.04$ & $-3.78\pm0.41$ & 0.93\\
    LMC & $9.72\pm1.39$ & $0.62\pm0.03$ & $-3.10\pm0.37$ & 0.80\\
    M31 & $7.24\pm0.93$ & $0.66\pm0.04$ & $-2.39\pm0.25$ & 0.80\\
	\enddata
\end{deluxetable*}

\begin{deluxetable*}{cccccccc}[ht] \label{tab:relations_sigma}
    \tablecaption{Fitted model parameters of $\sigma_{\mathrm{gran}}$ with stellar luminosity and effective temperature in  $\sigma_{\mathrm{gran}} \propto 10^{a_{\sigma}} L^{\beta_{\sigma}} T^{\gamma_{\sigma}}_\mathrm{eff}$.}

	\tablewidth{0pt}
	\tablehead{
		\colhead{Galaxy} & \colhead{$a_{\sigma}$} & \colhead{$\beta_{\sigma}$} & \colhead{$\gamma_{\sigma}$} & \colhead{Correlation} \\
	}
	\startdata
	SMC & $16.63\pm3.34$ & $0.45\pm0.11$ & $-4.78\pm0.95$ & 0.64\\
    LMC & $5.47\pm1.81$ & $0.87\pm0.04$ & $-2.22\pm0.48$ & 0.86\\
    M31 & $6.49\pm1.22$ & $0.50\pm0.05$ & $-1.97\pm0.32$ & 0.66\\
	\enddata
\end{deluxetable*}

\end{CJK*}
\end{document}